\documentclass{aa}

\usepackage[]          {fontenc}
\usepackage            {color}
\usepackage            {graphicx}
\usepackage            {amssymb}
\usepackage[authoryear]{natbib}
\usepackage            {6293macr}

\makeatletter

\providecommand{\boldsymbol    }[1]{\mbox{\boldmath $#1$}}

\def\vec{\bar}

\newcommand{\eg    }{{e.g.,\ }}
\newcommand{\ie    }{{i.e.,\ }}

\newcommand{\nnb   }{$\nu\bar\nu$}
\newcommand{\dif   }{{\rm d}}
\newcommand{\Msun  }{{{\rm M}_\odot}}
\newcommand{\kB    }{k_{\rm B}}
\newcommand{\ergsec}{\,erg\,s$^{-1}$}

\def\lsim{\mathrel{\rlap{\lower 4pt \hbox{\hskip 1pt $\sim$}}\raise 1pt\hbox {$<$}}}
\def\gsim{\mathrel{\rlap{\lower 4pt \hbox{\hskip 1pt $\sim$}}\raise 1pt\hbox {$>$}}}

\makeatother

\begin{document}
 \authorrunning{R.\,Birkl et\,al.\ }

 \titlerunning {Neutrino pair annihilation near accreting black holes}

 \title        {Neutrino pair annihilation near accreting, stellar-mass black holes}

 \author       {R.\,Birkl\inst{1}, M.A.\,Aloy\inst{1,2}, H.-Th.\,Janka\inst{1},
                and E.\, M\"{u}ller\inst{1}}

 \institute    {Max-Planck-Institut f\"{u}r Astrophysik, Postfach 1317, D-85741 Garching, Germany
                \and
                Departamento de Astronom\'{\i}a y Astrof\'{\i}sica,
                Universidad de Valencia, 46100 Burjassot, Spain}

 \mail         {rbirkl@mpa-garching.mpg.de}

 \date         {\today}

 \abstract     {We investigate the deposition of energy and momentum due to the annihilation of
                neutrinos ($\nu$) and antineutrinos ($\bar\nu$) in the vicinity of steady,
                axisymmetric accretion tori around stellar-mass black holes (BHs). This process is
                widely considered as an energy source for driving ultrarelativistic outflows with
                the potential to produce gamma-ray bursts.}
               {We analyze the influence of general relativistic (GR) effects in combination with
                different neutrinosphere properties on the \nnb-annihilation efficiency and spatial
                distribution of the energy deposition rate.}
               {Assuming axial symmetry, we numerically compute the annihilation rate 4-vector. For
                this purpose we construct the local neutrino distribution by ray-tracing neutrino
                trajectories in a Kerr space-time using null geodesics. We vary the value of the
                dimensionless specific angular momentum $a$ of the central BH, which provides the
                gravitational field in our models. We also study different shapes of the
                neutrinospheres, spheres, thin disks, and thick accretion tori, whose structure
                ranges from idealized tori to equilibrium non-selfgravitating matter distributions.
                Furthermore, we compute Newtonian models where the influence of the gravitational
                field on the annihilation process is neglected.}
               {Compared to Newtonian calculations, GR effects increase the total annihilation rate
                measured by an observer at infinity by a factor of two when the neutrinosphere is a
                thin disk, but the increase is only $\approx25\%$ for toroidal and spherical
                neutrinospheres. Comparing cases with similar luminosities, thin disk models yield
                the highest energy deposition rates by \nnb-annihilation, and spherical
                neutrinospheres the lowest ones, independent of whether GR effects are included.
                Increasing $a$ from 0 to 1 enhances the energy deposition rate measured by an
                observer at infinity by roughly a factor of 2 due to the change of the inner radius
                of the neutrinosphere. General relativity and rotation cause important differences
                in the spatial distribution of the energy deposition rate by \nnb-annihilation.}
               {}


 \maketitle

\section{Introduction}

 It is widely believed that systems powering gamma-ray bursts (GRB) could be newborn, stellar-mass
 black holes (BHs) accreting matter at hyper-critical rates (up to several solar masses per second)
 from a surrounding accretion disk with a mass of some hundredth of a solar mass up to possibly a
 solar mass (see, \eg \citealp{2005.Piran}). These central engines may form in a `collapsar' event
 where the core of a massive, rotating Wolf-Rayet star collapses to a BH and the accretion of the
 stellar envelope may eventually lead to a GRB-supernova event with relativistic mass ejection along
 the rotation axis \citep{1993.Woosley,1999.MacFadyen,2000.Aloy}. Accreting BHs may also be the
 remnants of mergers of two compact objects in close binaries \citep{1989.Eichler,1993.Mochkovitch}.
 In the first scenario the system is embedded in the envelope of the progenitor star, and the
 accretion disk is fed by stellar matter yielding very long ($\sim10-1000\,$s) accretion time scales
 comparable to the collapse time scale of the progenitor star. In the second scenario viscous
 transport in the accretion torus sets the secular time scale of the system ($\sim0.01-1\,$s) formed
 during the merger.

 The conditions in the vicinity of steady-state, hyperaccreting BHs have been analytically studied
 by \cite{1993.Jaroszynski,1996.Jaroszynski,1999.Popham,2001.Narayan,2002.Matteo,2002.Kohri,
 2006.Chen}, who determined the efficiency of energy loss by neutrino emission and the efficiency of
 energy conversion by neutrino-antineutrino (\nnb) pair annihilation into electrons and positrons.
 Three-dimensional hydrodynamic simulations have explored the {\em time-dependent} accretion in
 BH-torus systems, which are the remnants of neutron star--neutron star (NS+NS) and NS+BH mergers
 \citep{1999.Ruffert,2004.Setiawan,2004.Lee,2005.Lee.A,2005.Lee.B}. These investigations considered
 a rather compact torus (typical size: $\sim 10 - 20$ Schwarzschild radii) containing between a few
 hundredth of a solar mass and some $0.1\,\Msun$. Such torus masses result from NS+NS and NS+BH
 merger simulations \citep{1999.Janka,1999.Ruffert,2002.Janka,2003.Rosswog,2003.Shibata,
 2005.Shibata,2005.Oechslin}. The torus is partly opaque to neutrinos because of its high density.
 Typically, neutrino luminosities in excess of $10^{53}\,$erg$\,$s$^{-1}$ are produced. Under these
 conditions, the reactions $\nu + \bar\nu \rightarrow {\rm e}^+ + {\rm e}^- \rightarrow \gamma +
 \gamma$ give rise to energy deposition in the close vicinity of the BH at rates ranging from
 several $10^{49}\,$erg$\,$s$^{-1}$ up to more than $10^{51}\,$erg$\,$s$^{-1}$ \citep{1999.Ruffert,
 1999.Janka,2004.Setiawan,2005.Setiawan}. The resulting e$^+$e$^-$-pair plasma-photon fireball may
 power an ultrarelativistic outflow of baryons with typical Lorentz factors of $10^2 - 10^3$,
 provided the baryon loading (\ie the baryon rest mass compared to the internal energy in the
 outflow) remains sufficiently low (see, \eg \citealp{2005.Aloy} and references therein).

 Several studies involving different levels of sophistication have been performed, in order to
 determine the amount of energy which can be released by \nnb-annihilation near a (rotating)
 stellar-mass BH.

 \cite{1993.Jaroszynski,1996.Jaroszynski} investigated stationary configurations consisting of
 massive, dense and non-selfgravitating tori with different angular momentum distributions and
 different specific entropies, orbiting stellar-mass Kerr BHs. In particular, he considered the
 neutrino emission of isentropic tori in the external field of Kerr BHs as an approximation of what
 results from merger events or failed supernovae. Using the neutrino opacities of
 \cite{1986.Burrows} and assuming that the neutrinos have an equilibrium (Fermi-Dirac) distribution
 given by the temperature and chemical potential at the neutrinosphere, \cite{1996.Jaroszynski}
 determined the neutrino radiation field at a given point by following backwards null geodesics in
 Kerr spacetime until a point is reached in the torus where the neutrino optical depth is $\tau_{
 \nu} \approx 1$. He found that the energy deposition rate due to \nnb-annihilation increases with
 increasing entropy and with increasing spin of the Kerr BH. He also claimed that BH-torus
 configurations resulting from NS+NS merger events do not provide sufficient energy to likely be
 sources of GRBs.

 Extending earlier Newtonian calculations of \cite{1986.Cooperstein} and \cite{1987.Goodman},
 \cite{1999.Salmonson} analytically determined the proper energy deposition rate per unit proper
 time by \nnb-annihilation near the surface of a spherical NS (the neutrinosphere was assumed to
 coincide with the NS surface) including general relativistic (GR) effects. They concluded that the
 inclusion of the GR effects of ray bending and redshift (neutrino trajectories were computed in the
 Schwarzschild metric) enhances the proper deposition rate per unit of proper time compared to the
 Newtonian values by up to a factor of 4 for a neutrinosphere located at 2.5 Schwarzschild radii
 relevant for a proto-NS. An enhancement factor of 30 is possible, if the radius of the
 neutrinosphere shrinks to 1.5 Schwarzschild radii, which happens during the collapse of a NS to a
 BH.

 \cite{2000.Asano} studied the influence of GR effects on the \nnb-annihilation rate assuming two
 different geometries of the \nnb-emitting regions: a spherically symmetric emission region, and a
 thin disk emitter surrounding a Schwarzschild BH. The spherical geometry was already studied by
 \cite{1999.Salmonson}, but \cite{2000.Asano} improved on this work by approximately taking into
 account that some fraction of the deposited energy might not escape from the gravitational
 potential well, and thus cannot contribute to power GRBs. For disk-shaped neutrinospheres they
 determined the energy deposition rate per unit world time for an observer located at infinity by
 computing the annihilation rate near the symmetry axis. The latter restriction allows for a
 semi-analytic treatment by making use of the axial symmetry of the neutrino source. Contrary to
 \cite{1999.Salmonson}, they found that GR effects do not substantially change the energy deposition
 rate, neither for the spherically symmetric case nor for the disk case. This discrepancy arises,
 because the energy deposition rates were calculated with different constraints in both
 investigations. \cite{1999.Salmonson} considered the proper energy deposition rate per unit proper
 time, which is enhanced by the GR effects of gravitational blueshift and bending of trajectories,
 for a fixed neutrino luminosity at infinity, while \cite{2000.Asano} computed the (local) energy
 deposition rate per unit world time by assuming a given value of the effective temperature of the
 neutrinosphere and thus of the neutrinosphere luminosity. The inverse redshift factor relating the
 local luminosity to the luminosity at infinity explains the enhancement of the rate in case of
 \cite{1999.Salmonson}. However, as argued by \cite{2000.Asano}, the neutrinosphere luminosity,
 which is restricted or provided by models of the GRB engine, is the appropriate quantity to use
 when determining the influence of GR effects on the \nnb-annihilation process.

 Improving on their earlier work, \cite{2001.Asano} considered a non-selfgravitating, geometrically
 infinitely thin accretion disk surrounding a Kerr BH. The disk was either assumed to be isothermal
 or to have a prescribed temperature gradient. They found that if the energy deposition (along the
 rotation axis) is mainly due to neutrinos from the central part of the disk near the horizon,
 redshift effects are dominant, and the energy deposition rate is consequently reduced compared to
 the case, in which GR effects were neglected during the computation. Instead, if neutrinos that are
 emitted at larger radii dominate the energy deposition rate, GR bending effects become important.
 This causes a GR enhancement of the energy deposition rate by a factor of about 2, irrespective of
 the spin of the Kerr BH.

 \cite{2003.Miller} considered the full three-dimensional problem of numerically computing the
 \nnb-annihilation around a Kerr BH / thin accretion disk system using the full geodesic equations.
 They determined the energy-momentum deposition rate 4-vector per unit 4-volume in a given local
 observer's orthonormal frame. From this covariant quantity they derived the (coordinate-dependent,
 i.e. nonconvariant) energy-momentum deposition rate per proper time in Boyer-Lindquist coordinates
 for their observer, which is an adequate approximation to the 4-momentum per proper time
 at the observer's location as long as the observer is not too close to the horizon of the BH.
 \cite{2003.Miller} imaged the accretion disk for a specified, but arbitrary off-axis observer. They
 did this by integrating geodesics along the Boyer-Lindquist energy-momentum vectors until they
 either hit the BH, the disk, or reach $r=50M$. They confirmed the findings of \cite{2001.Asano} as
 to an only moderate GR enhancement of the energy deposition rate near the rotation axis, but they
 also showed that the dominant contribution to the energy deposition rate comes from near the
 surface of the disk, where the rate is a factor 10--20 times larger than on the rotation axis. This
 enhancement of the rate is independent of the spin of the BH. \cite{2003.Miller} performed their
 analysis for five Kerr BHs with different angular momenta, and for each case they binned the
 Boyer-Lindquist time component of the energy-momentum deposition rate 4-vector reaching $r=50M$
 into 20 polar angle bins. The resulting energy-momentum deposition rate per unit solid angle (as a
 function of polar angle) peaks along the surface of a cone centered on the rotation axis. The cone
 has a half-opening angle of $\pi/4$, because the spatial components of the energy-momentum
 deposition rate 4-vector are tilted by about 45 degrees towards the rotation axis near the surface
 of the disk. The total (integrated over polar angle) energy-momentum deposition rate approximately
 varies linearly with the spin of the BH, the energy deposition rate of a maximally rotating
 Kerr BH being about twice that of a non-rotating Schwarzschild BH. This is in rough agreement with
 the results of \cite{1993.Jaroszynski}.

 The previous studies discussed above were concerned with a restricted set of specific, idealized
 cases: spherical neutrinospheres (\citealp{1999.Salmonson,2000.Asano}), geometrically infinitely
 thin disks and hence neutrinospheres (\citealp{2000.Asano,2001.Asano,2003.Miller}), and
 specifically designed sequences of torus models without the possibility to identify the separate
 influence of different components and properties of the neutrino emitting system (\citealp{
 1993.Jaroszynski,1996.Jaroszynski}). In the following we present a more comprehensive and
 systematic study of energy-momentum deposition by \nnb-annihilation. We analyze how the neutrino
 distribution and the resulting \nnb-annihilation are influenced by GR effects, the geometry and the
 properties of the neutrinosphere, and the mass and spin of the central BH. To this end, we examine
 the most likely configurations occurring in compact astrophysical systems: spherical
 neutrinospheres present in NSs, and thin disks and thick tori surrounding stellar-mass BHs. We
 explicitly remark that the absolute values of the neutrino luminosity and of the energy-momentum
 deposition rate by \nnb-annihilation do not play an important role for the discussion in this
 paper. Due to the simplicity of the neutrino source models considered here, our numbers for these
 quantities should not be interpreted quantitatively. They sensitively depend on the location and
 temperature of the neutrinospheres, which must be expected to be different in detailed hydrodynamic
 simulations that include some treatment of the neutrino transport. However, for the purpose of
 comparing the different effects, on which this paper focuses, only the relative changes of the
 neutrino luminosity and \nnb-annihilation matter.

 The paper is organized as follows: In Sect.\,\ref{sec:theory} we discuss the theoretical
 fundamentals for calculating the \nnb-annihilation rate in a given Kerr spacetime, and how we
 construct equilibrium accretion tori surrounding Kerr BHs. In Sect.\,\ref{sec:numerics} we give a
 description of the numerical implementation of the ray-tracing algorithm used to calculate the
 annihilation rate. The results of our parameter study are discussed in Sect.\,\ref{sec:results},
 and the conclusions of our work are presented in Sect.\,\ref{sec:conclusions}. Finally, technical
 details concerning the calculation of the annihilation rate are given in App.\,
 \ref{sec:appendix_maths}, and convergence tests performed on our ray-tracing algorithm are
 presented in App.\,\ref{sec:appendix_convergence}.

\section{Theoretical fundamentals}
 \label{sec:theory}

 Unless stated otherwise, we use geometrized units throughout this paper, so that $c=G=1$, where
 $c$ is the speed of light in vacuum and $G$ is Newton's gravitational constant. Greek and Roman
 indices denote spacetime components (0--3) and spatial components (1--3) of 4-vectors,
 respectively. The signature of the metric is chosen to be $\left(+,-,-,-\right)$, and 3-vectors are
 denoted by a bar above the respective symbol, e.g. $\vec{A}$.

\subsection{Calculation of the annihilation rate}
 \label{sec:annihliation_rate}

 \begin{figure}
  \centerline{\includegraphics[width=0.5\textwidth]{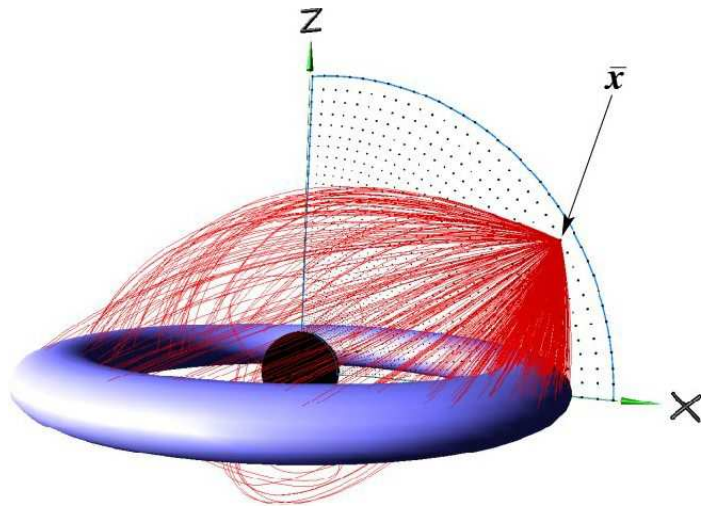}}
  \caption{Ray-tracing of neutrinos in a Kerr BH spacetime. At every point $\vec{x}$ where the
           annihilation rate is to be calculated, geodesics (red curves) arriving from random
           directions are traced back until they hit the neutrinosphere (blue torus). The
           computational grid is marked by the small black dots and its boundary by the blue circle.
           {\em See the electronic edition for a color version of the figure.}}
  \label{fig:raytracing}
 \end{figure}

 In order to compute the deposition of 4-momentum $Q_i^\alpha\left(t,\vec{x}\right)$ by the
 annihilation of neutrinos and antineutrinos of flavor $i$ with $i\in\left\{ {\rm e}, \mu,
 \tau\right\}$ into e$^+$e$^-$-pairs at a spatial point $\vec{x}$ per unit of time and unit of
 volume, we follow a formalism very close to that of \cite{2003.Miller}.

 In flat spacetime the local annihilation rate is computed from the Lorentz invariant neutrino and
 antineutrino phase space distribution functions $f_{\nu_i} = f_{\nu_i} \left(t,\vec{x},
 \vec{p}\right)$ and $f_{\bar{\nu}_i} = f_{\bar{\nu}_i} \left(t,\vec{x},\vec{p}'\right)$ (for their
 exact definition see Appendix\,\ref{sec:appendix_maths}) by the following integral
 \begin{equation}
  Q_i^\alpha=\int\dif^3p \dif^3p' A_i^\alpha\left(\vec{p},\vec{p}'\right)f_{\nu_i} f_{\bar\nu_i}\, ,
  \label{eq:1}
 \end{equation}
 where the $A_i^\alpha$ are functions defined in Appendix \ref{sec:appendix_maths}. They are based
 on a covariant generalization of the expressions given in \citet{1997.Ruffert}. Note that although
 we are concerned only with time-independent models here, we include the argument $t$ in the
 distribution functions $f_{\nu_i}$ and $f_{\bar{\nu}_i}$ for the sake of generality. In the
 following we omit the indices of $f$.

 Except for minor redefinitions Eq.\,(\ref{eq:1}) can also be used in curved spacetimes \citep{
 2003.Miller}, because the annihilation process, which is a microphysical phenomenon, happens so
 rapidly and on such small length scales that the effects of stellar gravitational fields can be
 safely neglected. Gravity affects only the propagation of the (anti)neutrinos between their
 emission (or emergence from the neutrinosphere) and their annihilation.

 In our models the gravitational field, which leads to ray bending and redshift, is provided by the
 central Kerr BH of mass $M$ and (dimensionless) angular momentum parameter $a \equiv J/M^2$ (where
 $J$ is the angular momentum of the BH, and $0 \le a \le 1$), whose metric is given in
 Boyer-Lindquist coordinates $\left(t, r, \theta, \phi \right)$ by
 \begin{equation}
  \dif s^2 =   g_{tt}\, \dif t^2 + g_{\phi\phi}\, \dif \phi^2
            + 2 g_{t\phi}\, \dif t \dif \phi
            + g_{rr}\, \dif r^2 + g_{\theta\theta}\, \dif \theta^2
  \label{kerrmetric}
 \end{equation}
 with
 \begin{eqnarray*}
  g_{tt}          &=& 1 - \frac{2 M r}{\rho^2}
  \, , \\
  g_{\phi\phi}    &=& - \left[r^2 + M^2 a^2 +
                            \frac{2 r M^3 a^2 \sin^2 \theta}{\rho^2}
                       \right] \sin^2 \theta
  \, , \\
  g_{t\phi}       &=& \frac{2 r M^2 a \sin^2 \theta}{\rho^2}
  \, , \\
  g_{rr}          &=& -\frac{\rho^2}{\Delta}
  \, , \\
  g_{\theta\theta}&=& -\rho^2
  \, ,
 \end{eqnarray*}
 where
 \begin{eqnarray*}
  \Delta & = & r^2 -2Mr + (aM)^2 \, , \\
  \rho^2 & = & r^2 + (aM)^2 \cos^2 \theta \, .
 \end{eqnarray*}

 Let us now consider an observer located at the point $\vec{x}$ where the annihilation happens
 (Fig.\,\ref{fig:raytracing}). We will refer to this observer as the local observer, and the
 quantities measured in his frame are denoted by the subscript `L'. The local frame is defined by an
 orthonormal base $\{\boldsymbol{e}_t, \boldsymbol{e}_r, \boldsymbol{e}_\theta, \boldsymbol{e}_\phi
 \}$ such that a local observer is at rest in the global $\left(r, \theta, \phi \right)$ coordinate
 system, i.e. its 4-velocity fulfills $u^a=0$ (the resulting orthonormal base vectors are given in
 App.\,\ref{sec:appendix_maths}). Hence, one cannot calculate the annihilation rate inside the
 ergosphere, where no observers can be at rest. This restriction could be lifted by using observers
 dragged along by the BH, i.e. for observers with $u_a = g_{a\beta}u^\beta=0$. However, the energy
 released inside the ergosphere will end up in the BH, and thus it is of no relevance for a GRB. We
 therefore exclude the region inside of the ergosphere from our analysis.

 In the immediate surroundings of a point $\vec{x}$ a curved spacetime can be approximated by a
 tangential flat spacetime. Thus, a local observer can use Eq.\,(\ref{eq:1}) to calculate the
 annihilation rate. Consequently, one has to generalize only the momenta $\vec{p}$ in
 Eq.\,(\ref{eq:1}) to $\vec{p}_{\rm L}$, in order to compute the annihilation rate $Q_i^\alpha$
 in the frame of a local observer.

 For doing so, one also needs to calculate the (anti)neutrino distribution functions in the frame of
 the local observer $f\left(t, \vec{x}, \vec{p}_{\rm L} \right)$. We take first the 4-momentum
 vector components $p_{\rm L}^\alpha = \left(E_{\rm L},\vec{p}_{\rm L}\right)$, with\footnote{In the
 local frame $\left|\vec{p}_L\right|^2 = \left(p_L^1\right)^2 + \left(p_L^2\right)^2 + \left(p_L^3
 \right)^2$.} $E_{\rm L} = \left|\vec{p}_{\rm L}\right|$, measured in the local frame and transform
 them into the components
 \[ p^\alpha=p_{\rm L}^\beta\boldsymbol{e}_{\beta}{}^\alpha\]
 measured in the global Boyer-Lindquist frame. Hence, the spatial components $\vec{p} = p^{a}$ are
 functions depending on $\vec{p}_{\rm L}$, such that we get a new distribution function $f\left(t,
 \vec{x}, \vec{p}\right) = f\left(t,\vec{x},\vec{p}_{\rm L} \left(\vec{p}\right)\right)$. Assuming
 now that the (anti)neutrinos do not experience collisions, their propagation is described by the
 collisionless Boltzmann equation in curved spacetime (see, \eg \citealt{1973.Misner})
 \begin{equation}
  \frac{ \dif f \left(t (\lambda), \vec{x} (\lambda),
         \vec{p} (\lambda) \right) }{  \dif \lambda} = 0 \, ,
  \label{eq:2}
 \end{equation}
 where $\lambda$ is a parameter along the neutrino path. This equation implies that the distribution
 function is constant along particle trajectories. Since we consider (anti)neutrinos to be massless,
 their propagation takes place along null geodesics. The latter are given by
 \begin{eqnarray*}
  \frac{\dif x^\alpha}{\dif\lambda} & = & p^{\alpha} \, , \\
  \frac{\dif p^\alpha}{\dif\lambda} & = &
        -\Gamma_{\beta\gamma}^\alpha p^\alpha p^\beta  \, ,
 \end{eqnarray*}
 with $g_{\alpha\beta}p^\alpha p^\beta=0$ (see, \eg \citealt{1998.Riffert}). To compute the
 distribution function $f\left(t, \vec{x}, \vec{p} \right)$ of either a neutrino or an antineutrino
 at the spacetime location $\left(t, \vec{x} \right)$ of momentum $\vec{p}$, we simply use the above
 equations to trace the path of that (anti)neutrino back in time until it hits its neutrinosphere
 (Fig.\,\ref{fig:raytracing}). In most cases this will never happen, and then $f\left(t, \vec{x},
 \vec{p} \right) = 0$. In the other cases, we get an emission event at $\left(t_{{\rm E}},
 \vec{x}_{\rm E}\right)$ with a neutrino 4-momentum $p_{\rm E}^\alpha = \left(E_{\rm E},
 \vec{p}_{\rm E}\right)$. Hence, using Eq.\,(\ref{eq:2}) this implies
 \begin{equation}
  f\left(t,\vec{x},\vec{p}\right) = f \left(t_{\rm E},\vec{x}_{\rm E},
                                            \vec{p}_{\rm E}\right) \, .
  \label{eq:3}
 \end{equation}
 Note that $f\left(t_{\rm E},\vec{x}_{\rm E},\vec{p}_{\rm E}\right)$ is the distribution function at
 the emission point in the global frame, and it has to be related to the corresponding function in
 the frame comoving with the neutrinosphere. This function has the form of a black body for
 fermions (the chemical potential is neglected, which is justified for the typical conditions
 met in the systems we are interested in), and therefore it depends only on the comoving frame
 energy $E_{\rm C}$. Taking into account that the 4-velocity $u_{\rm C}\left(t,\vec{x}_{\rm E}
 \right)$ of the neutrinosphere is equal to the time-like base vector of the comoving frame, the
 comoving frame energy $E_{\rm C}$ is the projection of the 4-momentum $p_{\rm E}$ onto $u_{\rm C}
 \left(t,\vec{x}_{\rm E}\right)$:
 \[ E_{\rm C}=p_{\rm E} \cdot u_{\rm C}  \, . \]
 Hence, the rhs of Eq.\,(\ref{eq:3}) can be expressed in terms of the comoving frame distribution
 function as
 \[
  f \left(t_{\rm E}, \vec{x}_{\rm E}, \vec{p}_{\rm E} \right) =
  \frac{1}{1 + \exp \left( \frac{E_{\rm C}}{\kB T_{\rm C}} \right) }  \, ,
 \]
 where $\kB = 1.381 \cdot 10^{-16}\,$erg\,K$^{-1}$ is the Boltzmann constant, and $T_{\rm C}$ the
 temperature of the neutrinosphere.

\subsection{Neutrinospheres}
 \label{sub:theory_neutrinosphere}

 The conditions at the neutrinosphere are of importance when calculating the neutrino and
 antineutrino luminosities and \nnb-annihilation, because the neutrino emission properties
 sensitively depend on the neutrinosphere temperature $T_{\rm C}$, and on the radiating area. We
 therefore investigate various neutrinosphere geometries, constructed in two different ways. In the
 first approach, we prescribe the location and the shape of the neutrinosphere for several idealized
 BH-accretion disk systems and study their influence on the annihilation rate. In the second, more
 elaborate approach we proceed similar to \cite{1993.Jaroszynski,1996.Jaroszynski}, and model the
 accretion disk as a non-selfgravitating, stationary, geometrically thick equilibrium torus rotating
 around a Kerr BH with mass $M$ and rotation parameter $a$.

 The distribution of the specific enthalpy of the equilibrium tori is computed according to the
 method of \cite{1978.Abramowicz} for disks with uniform specific angular momentum ($l$). Our
 equilibrium torus models are constructed by fixing their mass ($m_\mathrm{tor}$) and the value of
 the inner equatorial radius of the torus (see, e.g., \citealp{2002.Font}), which we choose to be
 $r_\mathrm{in}=4.1\,M$ and whose specification replaces the specification of $l$. Once the specific
 enthalpy distribution is found, we obtain the rest mass density ($\rho$) distribution using the
 assumption that the equation of state is barotropic. We further assume that the dimensionless
 entropy per baryon carried by photons, $s_\gamma = 4 a_\gamma (k_{\rm B} T)^3/(3\rho/m_{\rm u})$,
 is given and uniform, too
 \footnote{The constants used in the definition of the photon entropy
  $s_\gamma$ are $a_\gamma = 8 \pi^5 /(15 h^3 c^3)= 2.082\cdot 10^{49}\,$erg$^{-3}$\,cm$^{-3}$,
  $m_{\rm u}=1.661\cdot 10^{-24}\,$g, and $h=6.626 \cdot10^{-27}\,$erg\,s.}.
 This assumption translates into a relation between the density and the temperature ($\rho \propto
 T^3$), which allows us to compute the latter. The resulting tori are radiation dominated for values
 of $s_\gamma \gsim 1$. In Sect.\,\ref{sub:results_equilibrium} we take $s_\gamma = 1$ in the tori,
 which corresponds to total entropies per baryon in the range of 5 to 12 and agrees with the values
 resulting from detailed hydrodynamic Newtonian simulations of mergers of compact binaries
 (\citealp{1999.Ruffert}; \citealp{2005.Setiawan}). We further assume that matter in the torus
 consists of free protons, neutrons, e$^-$, e$^+$, and photons. For the typical conditions in the
 torus, it is justified to neglect the effects of strong interactions on the equation of state. With
 the baryon rest-mass density $\rho$, electron fraction $Y_{\rm e} \equiv (n_{{\rm e}^-} - n_{{\rm
 e}^+})/n_{\rm B}$ (where $n_{\rm B}$, and $n_{{\rm e}^{\pm}}$ are the number densities of baryons,
 and of electrons and positrons, respectively) and $s_\gamma$ given, the thermodynamic state is
 unambiguously defined. The specification of $Y_{\rm e}$ is needed to determine the composition of
 the torus gas such that the neutrino opacities can be calculated, and its value is chosen to be
 $Y_{\rm e}=0.1$ in our paper. Therefore, an equilibrium torus is fully specified by the six
 parameters $M$, $a$, $m_\mathrm{tor}$, $s_\gamma$, $r_\mathrm{in}$, and $Y_{\rm e}$.

 For the equilibrium torus models the neutrino and antineutrino opacities are computed as in
 \cite{2001.Janka}. The neutrino spectrum is assumed to be that of a black body for fermions with
 zero chemical potential, and the neutrinospheres are determined as surfaces surrounding the BH, at
 which the optical depth for equilibration of (anti)neutrinos is $\tau = 2/3$. The integration to
 calculate $\tau$ is done along rays parallel to the symmetry axis, assuming neutrinos to have an
 average energy according to a thermal spectrum with the local temperature. We point out that this
 approach to estimate the neutrinospheric surfaces differs from that of \cite{1996.Jaroszynski}, who
 integrated neutrino trajectories back to the points where $\tau=1$. Our method nevertheless leads
 to neutrinosphere geometries, which are very similar to those of \cite{1996.Jaroszynski}.

 However, the neutrinosphere concept is only an approximation. Given some matter distribution, e.g.
 in an accretion torus, the neutrinospheres roughly separate optically thin from optically thick
 regions. Strictly speaking, there exists no well defined surface where neutrinos decouple
 instantaneously from the background, because the neutrino opacities depend on the neutrino energy.
 It is also a crude simplification to assume that there exists a surface, exterior to which
 neutrinos stream freely, and interior to which they are in equilibrium with matter and diffuse out
 suffering multiple scattering. Therefore our definition of the neutrinospheres appears sufficiently
 good to discuss the basic features of \nnb-annihilation in the vicinity of accreting tori in
 rotational equilibrium around BHs, although it is very approximative and does not take into account
 the local conditions along the curved geodesical paths of neutrinos.

\section{Numerical implementation}
 \label{sec:numerics}

 The \nnb-annihilation rate is computed on a spherical grid consisting of $N_{\rm r}$ points in
 radial and $N_\theta$ points in polar ($\theta$) direction, respectively. Typically, we use
 $N_\theta=100$ uniformly distributed points excluding a small conical region around the rotation
 axis of the BH for numerical reasons. The radial grid is non-equidistant, stretching from the
 BH-horizon to a radius slightly larger than the most distant radial point of the neutrinosphere
 (Figs.\,\ref{fig:field_idealized1}, \ref{fig:field_idealized2}, and \ref{fig:field_equilibrium}).
 The radial grid points are distributed according to $r_{i+1} = \left( 1 + \Delta\theta \right)
 r_{i}$, where $\Delta\theta$ is the angular grid spacing, and $i=1, \ldots, N_{\rm r}$ (typically
 $N_{\rm r}=100$). We assume axial and equatorial symmetry.

 For each of our models we trace $N_{\rm rays}=20000$ rays back in time in random directions,
 starting from every grid point of the spherical grid exterior to the neutrinospheres. The procedure
 is performed independently for neutrinos and antineutrinos (except when the neutrinospheres are
 identical). The time integration is performed with a fourth-order adaptive step-size Runge-Kutta
 method until the neutrino trajectory hits the neutrinosphere. To determine when a neutrinosphere is
 hit, we have developed the following mesh refinement algorithm. On a coarse grid of $150\times150$
 zones we label the zones belonging to the optically thick region (interior to the neutrinosphere),
 and we flag the zones which are crossed by the neutrinosphere (Fig.\,\ref{fig:raytracing}). The
 latter zones are further refined, such that each of them is covered by $150 \times 150$ fine zones.
 Again, we label each of the fine zones belonging to the interior of the neutrinosphere. The whole
 process has to be performed only once (before calculating the annihilation rate on the spherical
 grid). This algorithm allows one at each step of the ray-tracing to very rapidly check whether the
 optically thick region has been reached. In that case the last twenty steps of the ray-tracing are
 redone with improved accuracy. The described method can deal with arbitrary (axisymmetric)
 neutrinosphere geometries, but it is very inefficient when applied to cases involving {\em
 infinitely} thin disks. For these cases our algorithm branches to a different `hit-detection'
 procedure, and checks whether the $(x,y)$-plane (\ie the equatorial plane), containing the
 neutrinosphere and the thin disk, is intersected between the inner and outer radial edge of the
 disk. Most ray-traced neutrinos never hit the neutrinosphere, because either they are trapped by
 the BH or because they escape to infinity.

\section{Simulation results}
 \label{sec:results}

 Though we included all three (anti)neutrino flavors in the previous theoretical discussion, we
 consider the annihilation of only $\nu_{\rm e}$ and $\bar\nu_{\rm e}$ in our simulations, since the
 energy deposition is dominated by the latter process. The reasons for that are the reduced
 production and emission of heavy-lepton neutrinos by GRB accretion tori (see, \eg
 \citealp{1999.Ruffert}) and the fact that the weak coupling constants for \nnb-annihilation of muon
 and tau neutrinos are roughly a factor of five lower than those of electron-type neutrinos. Thus,
 in the following we use the shorthand notation $Q^\alpha$ for the annihilation rate $Q_{{\rm e}}^
 {\alpha}$ measured in the frame of a local observer.

 In order to quantify the total amount of energy released by \nnb-annihilation per unit of time, we
 consider four energy deposition rates, which are all based on volume integrals of the energy
 component $Q^t$ of the annihilation rate 4-vector. The first two rates are integrals of the rates
 measured in the local frames:
 \begin{eqnarray}
  \dot{E}_{\nu\bar\nu}^{\rm tot} & = &
  \int_{{\rm V_{\rm tot}}} \dif r\, \dif \theta\, \dif \phi\,
                           \sqrt{ -\det \left( g_{ab} \right) }\, Q^t
  \, , \label{edotltot}\\
  \dot{E}_{\nu\bar\nu}^{\rm up} & = &
  \int_{{\rm V_{\rm up}}}  \dif r\, \dif \theta\, \dif \phi\,
                           \sqrt{ -\det \left( g_{ab} \right) }\, Q^t
  \, . \label{edotlup}
 \end{eqnarray}
 Here $V_{\rm tot}$ and $V_{\rm up}$ are the volume of the whole computational grid and the volume
 of that part of the computational grid where $Q^r>0$, respectively. In the `up' region the radial
 component of the momentum vector of the e${^+}$e${^-}$-pair created by \nnb-annihilation directs
 outward. Ignoring hydrodynamic effects of the produced e${^+}$e${^-}$-photon plasma (i.e., adopting
 a `free-particle picture'), the energy deposited in this `up' region is likely to eventually reach
 a distant observer, i.e. it is the crucial region for our study. In contrast, the energy released
 between the event horizon and $V_{\rm up}$ is probably swallowed by the BH.

 We further consider the corresponding energy deposition rates measured by an infinitely distant
 observer (\citealp{1993.Jaroszynski}):
 \begin{eqnarray}
  \dot{E}_{\nu\bar\nu}^{{\rm tot},\infty} & = &
  \int_{{\rm V_{\rm tot}}} \dif r\, \dif \theta\, \dif \phi\,
                           \sqrt{ -\det \left( g_{\alpha\beta} \right) }\, Q^t
  \, , \label{edotitot} \\
  \dot{E}_{\nu\bar\nu}^{{\rm up},\infty} & = &
  \int_{{\rm V_{\rm up}}}  \dif r\, \dif \theta\, \dif \phi\,
                           \sqrt{ -\det \left( g_{\alpha\beta} \right) }\, Q^t
  \, . \label{edotiup}
 \end{eqnarray}
 Note that while in Eqs.\,(\ref{edotltot}) and (\ref{edotlup}) the integrands contain the
 determinant of the 3-metric (Latin indices), Eqs.\,(\ref{edotitot}) and (\ref{edotiup}) involve the
 determinant of the 4-metric (Greek indices). The latter two energy deposition rates are generally
 smaller than the corresponding rate computed as the integral of $Q^t$ in the local frames, because
 they include the redshift due to the gravitational field. If GRBs are fueled by the process of
 \nnb-annihilation, $\dot{E}_{\nu \bar\nu}^{{\rm up}, \infty}$ (Eq.\,\ref{edotiup}) represents a
 rough estimate for the maximum luminosity of the GRB event.

 Two other quantities relevant for analyzing our models are the `local luminosity' radiated from
 the neutrinosphere,
 \begin{equation}
  L_{\nu} = \frac{\pi \kB^4}{h^3}\, F_3(0)
            \int_{\nu-{\rm sphere}} \dif \sigma\, T_{\rm C}^4 \, ,
  \label{eq:Lnu}
 \end{equation}
 where we assume blackbody emission from an isotropically radiating surface, and the corresponding
 `luminosity' for an observer at infinity,
 \begin{equation}
  L^{\infty}_{\nu} = \frac{\pi \kB^4}{h^3}\, F_3(0)
            \int_{\nu-{\rm sphere}} \dif \sigma\, g_{tt} T_{\rm C}^4 \, .
  \label{eq:Lnui}
 \end{equation}
 Here $F_3(0) = \int_0^\infty \dif x \left\lbrack x^3 / (e^x+1) \right\rbrack$ and $\dif \sigma$ is
 the appropriate GR surface element of the neutrinosphere. The factor $g_{tt}$ in Eq.\,
 (\ref{eq:Lnui}) accounts for the gravitational redshift. Note that what we call `luminosity' here
 is actually the total energy emission rate from the neutrinosphere, which is not the quantity
 directly measured by an observer at a fixed position, because we ignore the direction-dependent
 variation of the surface area of the neutrinosphere visible to an observer.

 The evaluation of the quantities $\dot{E}_{\nu\bar\nu}^{{\rm tot},\infty}$ and $\dot{E}_{\nu\bar
 \nu}^{{\rm up},\infty}$ (see Eqs.\,(\ref{edotitot}) and (\ref{edotiup})) requires two steps. In the
 first step $Q^t$ is calculated at the annihilation point as described in Sect.\,
 \ref{sec:annihliation_rate}, including aberration effects due to the neutrinosphere rotation (for
 models with $l\neq0$). Using the obtained $Q^t$, Eqs.\,(\ref{edotitot}) and (\ref{edotiup}) are
 evaluated in the second step. Note that these two equations involve an approximation, because the
 spatial components $Q^a$ of the annihilation rate 4-vector do not enter. This approximation means
 that the momentum of the \nnb-annihilation plasma at the annihilation point is ignored in the
 transformation of the energy to infinity. In contrast to the evaluation of $Q^t$, neutrinosphere
 aberration effects are neglected in Eq.\,(\ref{eq:Lnui}), because this enormously simplifies the
 evaluation and can be justified by the use of an approximation in Eqs.\,(\ref{edotitot}) and
 (\ref{edotiup}) as explained above.

\subsection{Idealized models}
 \label{sub:results_idealized}

 \begin{table*}
  \tabcolsep=0.5mm
  \caption{\setlength{\baselineskip}{11pt} 
           Some properties of the simulated idealized models, whose isothermal neutrinosphere and
           antineutrinosphere are assumed to coincide and to have the same temperature. The
           quantities given in the columns of the table from left to right are: the mass $M$ and the
           (dimensionless) angular momentum parameter $a$ of the BH, the geometry of the
           (anti)neutrinosphere, the inner and outer radius (disk/torus) or just the radius (sphere)
           of the neutrinosphere; for toroidal neutrinospheres the ratio of the radial to the
           vertical diameter of its elliptical meridional cross section is given, too. The other
           quantities are the Lagrangian angular momentum $l$, measured in terms of the BH mass
           $M_{\rm GR}$ of the GR model, i.e. in case of a GR model we have $M_{\rm GR}=M\neq0$,
           whereas in case of a Newtonian model ($M=0$) the value of the Lagrangian angular momentum
           of the corresponding GR model is used (cgs values of $l$ are obtained by applying the
           conversion factor $\left(M_{\rm GR}/M_{\odot}\right)\cdot4.42\cdot10^{15}\,{\rm cm}^2\,{
           \rm s}^{-1}$), the comoving frame temperature $T_{{\rm C}}$ of the neutrinosphere (in the
           special model TEMP there is no emission outside of a cylinder around the system axis with
           radius $8M$), the total neutrinosphere luminosity $L^{\infty}_{\nu} = L^{\infty}_{\nu_{e}
           } + L^{\infty}_{\bar{\nu}_{e}}$ for an observer at infinity, the total energy deposition
           rate, $\dot{E}_{\nu \bar{\nu}}^{{\rm tot}}$, as the integral over the total volume of the
           local energy deposition rate $Q^{t}$, the total energy deposition rate $\dot{E}_{\nu \bar
           {\nu}}^{{\rm up}}$ in the `up' region (defined by $Q^{r}>0$), these total energy
           deposition rates measured by an infinitely distant observer, $\dot{E}_{\nu\bar{\nu}}^{{
           \rm tot}, \infty}$ and $\dot{E}_{\nu\bar{\nu}}^{{\rm up},\infty}$, and the efficiencies
           $q_{\nu \bar\nu}^{{\rm tot},\infty}\equiv\dot{E}_{\nu\bar\nu}^{{\rm tot}, \infty} / L^{
           \infty}_\nu$ and $q_{\nu \bar\nu}^{{\rm up},\infty} \equiv\dot{E}_{\nu\bar\nu}^{{\rm up},
           \infty} / L^{\infty}_\nu$, respectively.}
  \label{tab:idealized}
  \begin{center}
    \begin{tabular}{c|cc|cccc|c|cccccc}
	  \hline\hline  &&&&&&&&&&&&& \\*[-0.3cm]
     model & $M$ & $a$ & geometry & radii & $l$ & $T_{\rm C}$ & $L^{\infty}_{\nu}$ &
     $\dot{E}_{\nu\bar{\nu}}^{{\rm tot}}$ &
     $\dot{E}_{\nu\bar{\nu}}^{{\rm up}}$ &
     $\dot{E}_{\nu\bar{\nu}}^{{\rm tot},\infty}$ &
     $\dot{E}_{\nu\bar{\nu}}^{{\rm up},\infty}$ &
     $q_{\nu\bar{\nu}}^{{\rm tot},\infty}$ &
     $q_{\nu\bar{\nu}}^{{\rm up},\infty}$ \\
     name & $M_{\odot}$ &&& $M$ & $M_{\rm GR}$ & $10^{10}\rm K$ &
     $10^{52}$\ergsec &
     $10^{49}$\ergsec &
     $10^{49}$\ergsec &
     $10^{49}$\ergsec &
     $10^{49}$\ergsec &
     $10^{-3}$ & $10^{-3}$ \\*[0.06cm]
    \hline  &&&&&&&&&&&&& \\*[-0.35cm]
     D      &  2     &  0     & disk   & $6\leftrightarrow7.7$         & 0    & 5   & 0.33 & 0.54  &
               0.27  &  0.36  & 0.23   &  1.1  & 0.70 \\
     DN     &  0     &  0     & disk   & $6\leftrightarrow7.7$         & 0    & 5   & 0.39 & 0.15  &
               0.12  &  0.15  & 0.12   &  0.38 & 0.31 \\
     T      &  2     &  0     & torus  & $6\leftrightarrow7.1;\,1$     & 0    & 5   & 0.30 & 0.66  &
               0.13  &  0.44  & 0.11   &  1.5  & 0.37 \\
     TN     &  0     &  0     & torus  & $6\leftrightarrow7.1;\,1$     & 0    & 5   & 0.39 & 0.16  &
               0.095 &  0.16  & 0.095  &  0.41 & 0.24 \\
     S      &  2     &  0     & sphere &  3.4                          & 0    & 5   & 0.16 & 0.12  &
               0.12  &  0.083 & 0.083  &  0.52 & 0.52 \\
     SN     &  0     &  0     & sphere &  3.4                          & 0    & 5   & 0.39 & 0.066 &
               0.066 &  0.066 & 0.066  &  0.17 & 0.17 \\
    \hline
     DA1    &  2     &  1     & disk   & $6\leftrightarrow7.7$         & 0    & 5   & 0.33 & 0.56  &
               0.27  &  0.38  & 0.23   &  1.2  & 0.70 \\
     DL6.5  &  2     &  0     & disk   & $6\leftrightarrow7.7$         & 6.5  & 5   & 0.33 & 0.30  &
               0.28  &  0.25  & 0.24   &  0.76 & 0.73 \\
     DL4    &  2     &  0     & disk   & $6\leftrightarrow7.7$         & 4    & 5   & 0.33 & 0.38  &
               0.24  &  0.28  & 0.21   &  0.85 & 0.62 \\
     TL4    &  2     &  0     & torus  & $6\leftrightarrow7.1;\,1$     & 4    & 5   & 0.30 & 0.38  &
               0.12  &  0.23  & 0.099  &  0.78 & 0.33 \\
     DL4N   &  0     &  0     & disk   & $6\leftrightarrow7.7$         & 4    & 5   & 0.39 & 0.13  &
               0.11  &  0.13  & 0.11   &  0.32 & 0.29 \\
     TL4N   &  0     &  0     & torus  & $6\leftrightarrow7.1;\,1$     & 4    & 5   & 0.39 & 0.12  &
               0.071 &  0.12  & 0.071  &  0.31 & 0.18 \\
    \hline
     REF    &  3     &  0     & disk   & $6\leftrightarrow10$          & 0    & 5   & 2.1  & 8.7   &
               5.1   &  6.2   & 4.4    &  2.9  & 2.1  \\
     A2B    &  3     &  0     & torus  & $6.3\leftrightarrow9.7;\,2$   & 0    & 5   & 2.0  & 8.0   &
               3.1   &  5.0   & 2.7    &  2.4  & 1.3  \\
     AB     &  3     &  0     & torus  & $6.7\leftrightarrow9.3;\,1$   & 0    & 5   & 2.0  & 10    &
               2.3   &  5.7   & 2.0    &  2.9  & 1.0  \\
     A.5B   &  3     &  0     & torus  & $7.2\leftrightarrow8.8;\,0.5$ & 0    & 5   & 1.9  & 15    &
               2.0   &  7.6   & 1.7    &  4.0  & 0.89 \\
     SM3    &  3     &  0     & sphere & 5.7                           & 0    & 5   & 1.6  & 1.3   &
               1.3   &  1.1   & 1.1    &  0.70 & 0.70 \\
     RI4.5  &  3     &  0     & disk   & $4.5\leftrightarrow9.2$       & 0    & 5   & 2.1  & 13    &
               6.4   &  8.3   & 5.4    &  4.0  & 2.6  \\
     RI3    &  3     &  0     & disk   & $3\leftrightarrow8.5$         & 0    & 5   & 2.0  & 20    &
               6.8   & 11     & 5.5    &  5.6  & 2.8  \\
     TEMP   &  3     &  0     & torus  & $6.7\leftrightarrow9.3;\,1$   & 0    & 6.1 & 2.0  & 53    &
               6.9   & 27     & 5.7    & 14    & 2.9  \\
    \hline
     A.5    &  3     &  0.5   & disk   & $6\leftrightarrow10$          & 0    & 5   & 2.1  & 8.9   &
               5.1   &  6.3   & 4.4    &  3.0  & 2.1  \\
     A1     &  3     &  1     & disk   & $6\leftrightarrow10$          & 0    & 5   & 2.1  & 8.9   &
               5.1   &  6.5   & 4.4    &  3.1  & 2.1  \\
     L2.5   &  3     &  0     & disk   & $6\leftrightarrow10$          & 2.5  & 5   & 2.1  & 7.7   &
               5.0   &  5.7   & 4.2    &  2.7  & 2.0  \\
     L4     &  3     &  0     & disk   & $6\leftrightarrow10$          & 4    & 5   & 2.1  & 6.6   &
               4.9   &  5.1   & 4.2    &  2.4  & 2.0  \\
     L5     &  3     &  0     & disk   & $6\leftrightarrow10$          & 5    & 5   & 2.1  & 5.9   &
               4.9   &  4.8   & 4.2    &  2.3  & 2.0  \\*[0.05cm]
    \hline
   \end{tabular}
  \end{center}
 \end{table*}

 \begin{figure*}
  \vspace{0.19cm}
  \begin{center}
    \includegraphics[width=0.4\textwidth]{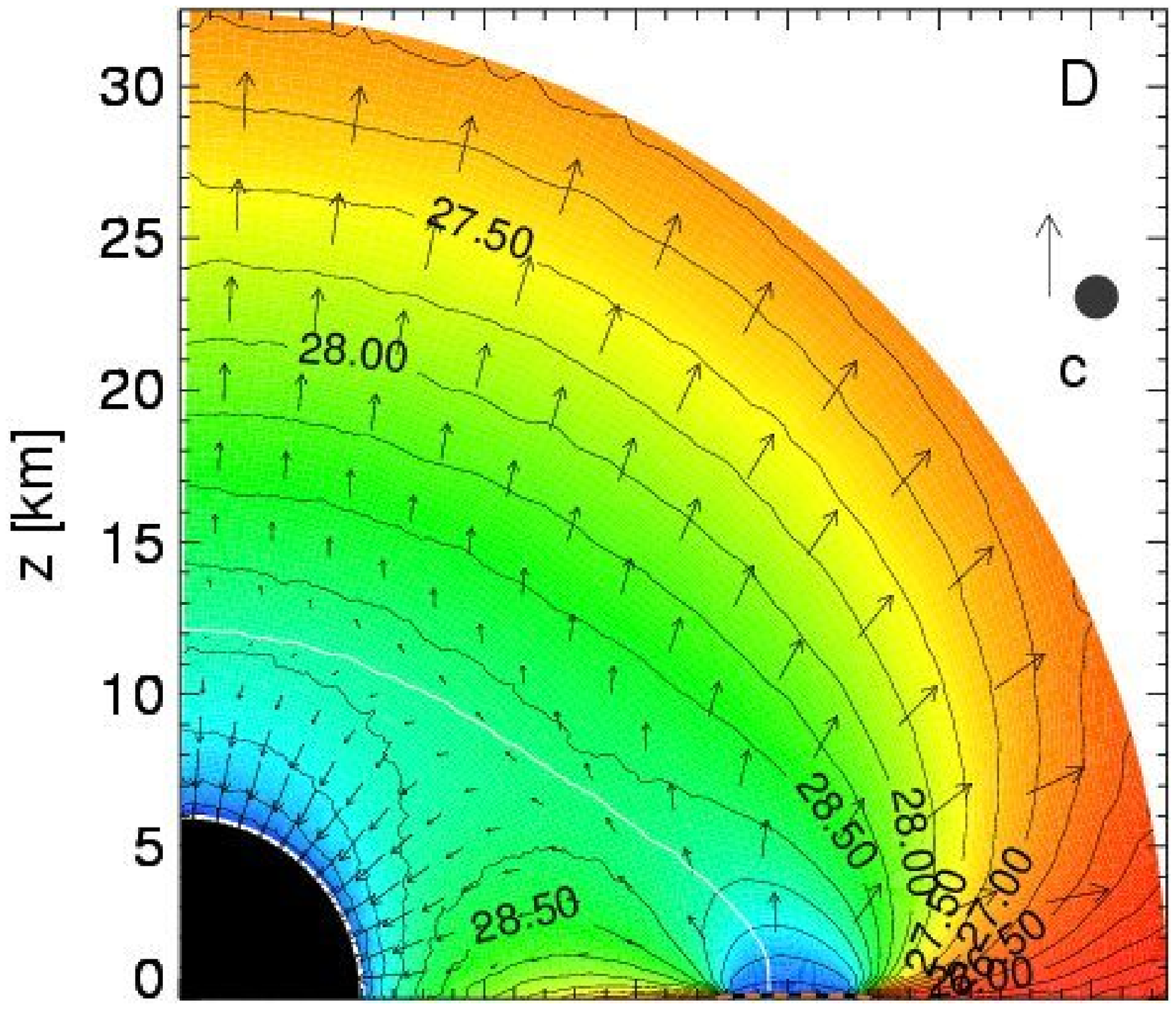}
   \hspace{1.75cm}
    \includegraphics[width=0.4\textwidth]{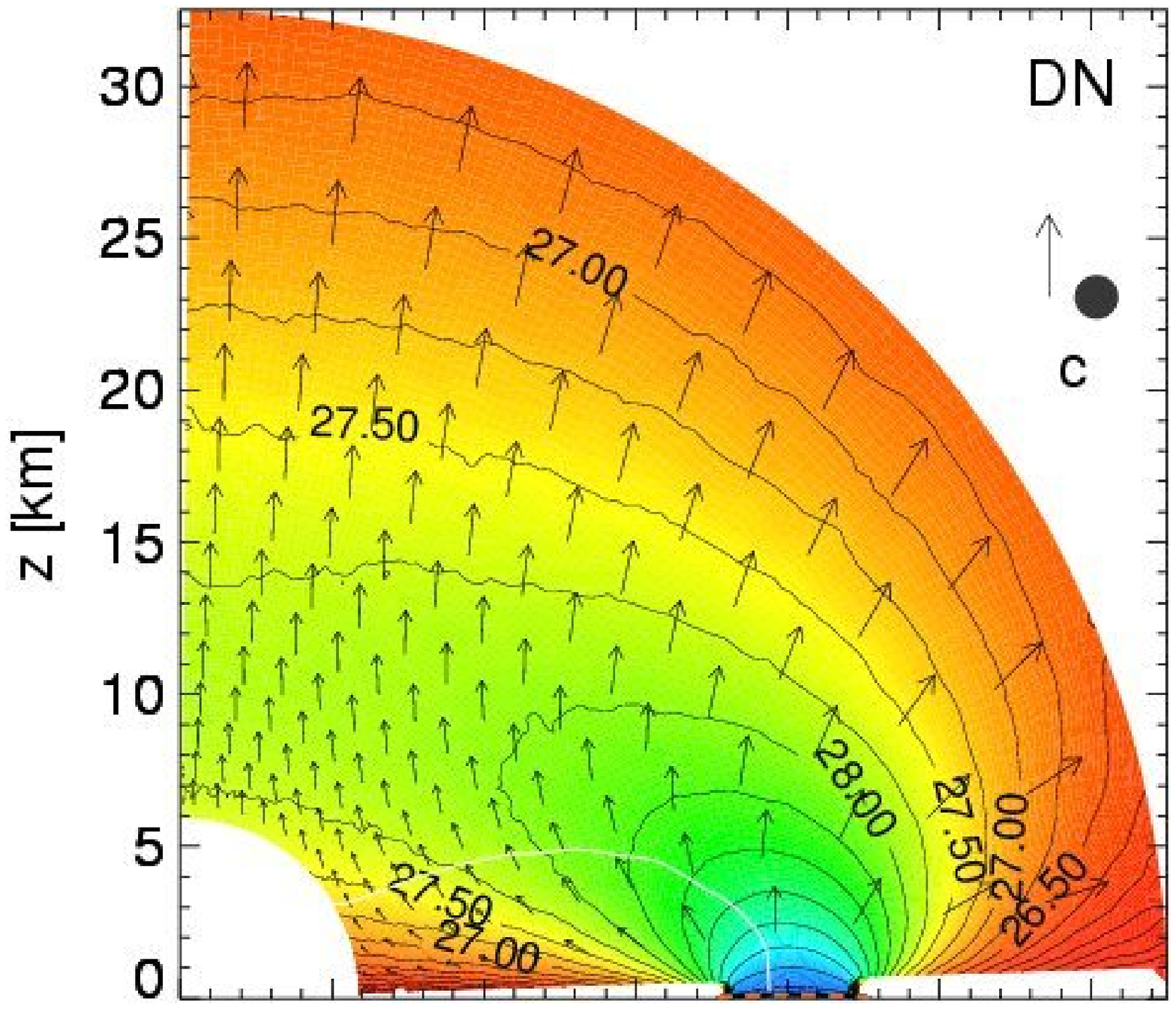}\\
   \vspace{0.6cm}
    \includegraphics[width=0.4\textwidth]{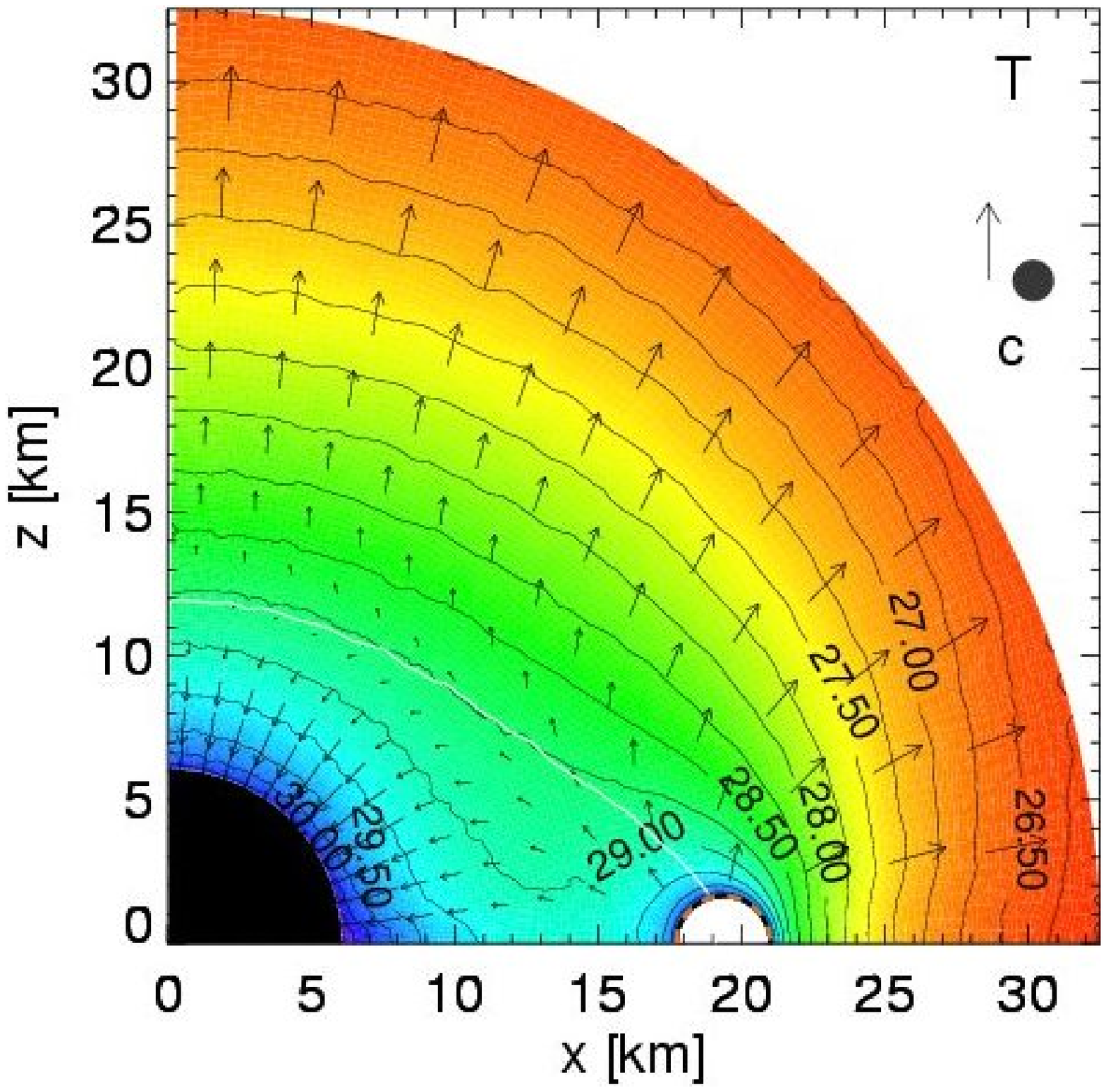}
   \hspace{1.75cm}
    \includegraphics[width=0.4\textwidth]{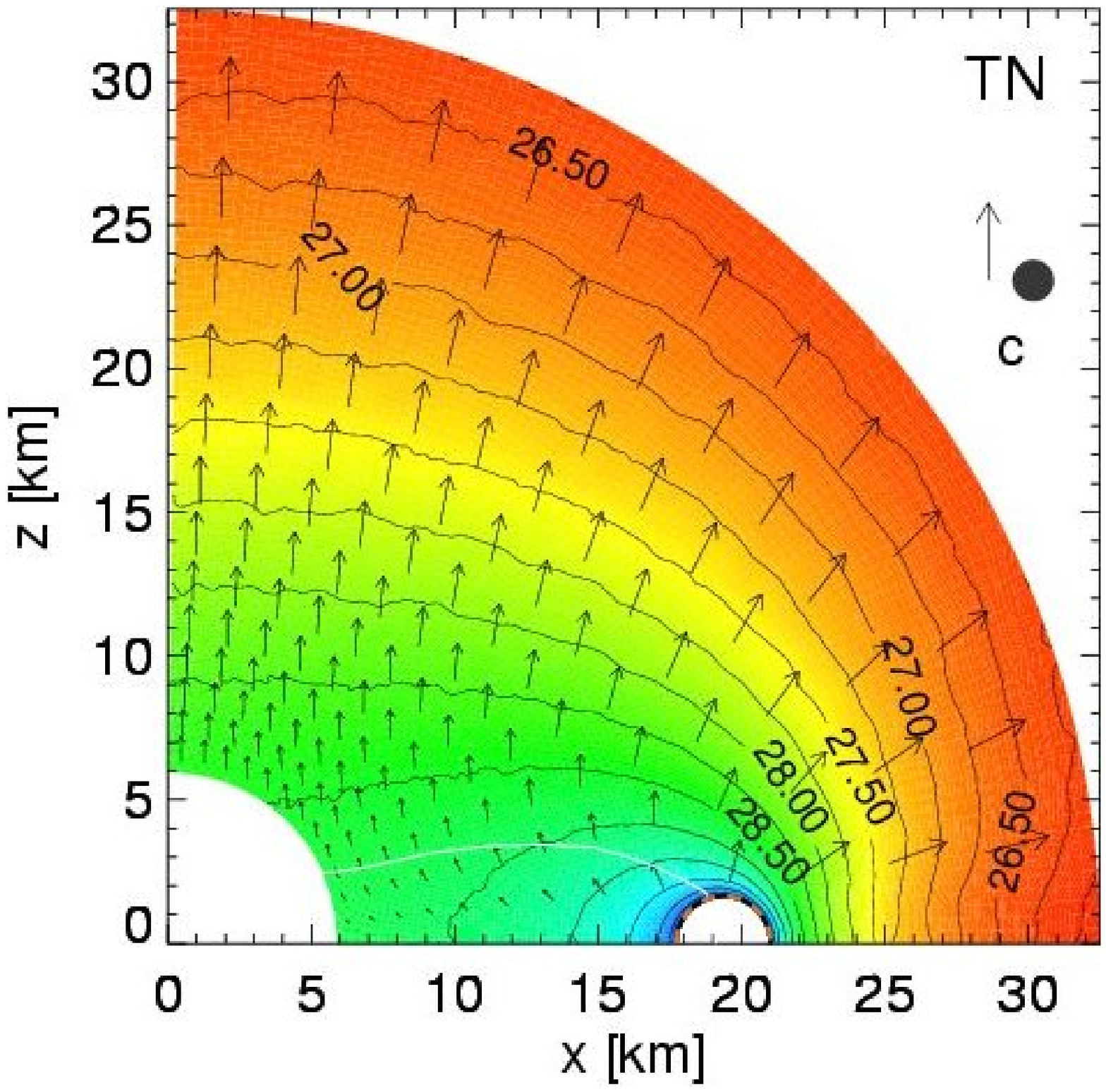}\\
   \vspace{0.3cm}
    \includegraphics{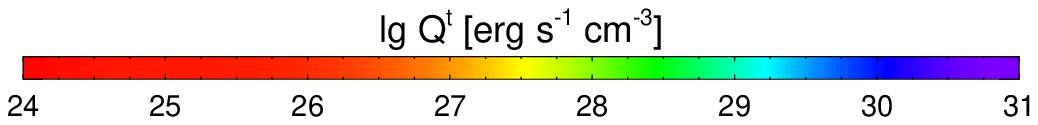}
  \end{center}
  \caption{Annihilation rate 4-vector $Q^{\alpha}$ measured in the frame of a local observer for
           models with idealized neutrinospheres (see Table \ref{tab:idealized}). The location of
           the neutrinosphere, which coincides with the antineutrinosphere, is marked by a dashed
           line. The value of the energy component $Q^{t}$ is color-coded, while the spatial vector
           $\overline{Q}$ is visualized by showing the spatial velocity vector $\vec{v}\equiv
           \overline{Q}/Q^{t}$, whose component coplanar (perpendicular) to the displayed
           $x-z$-plane is given by the black arrows (filled circle). The length (area) of the arrow
           (filled circle) is a linear measure of the size of the component, the maximum value
           ($c=1$) being represented by the arrow (filled circle) in the right upper corner of each
           panel. Note that for the neutrino evaluation the neutrinosphere is assumed to be
           non-rotating. Therefore the perpendicular component of $\vec{v}$ is zero here. The big
           black circular region in the left lower corner of the left two panels represents the BH.
           In the right two panels the corresponding region is chosen to be white, because these
           panels show Newtonian models where the GR effects of the BH were disregarded. Finally, in
           every panel a solid white line separates regions with positive $v^{r}$ from those with
           negative $v^{r}$. Note that in model DN the white region along the equatorial plane
           (x-axis) is caused by the saturation of the color scale due to a very low energy
           deposition rate.
           {\em See the electronic edition for a color version of the figure.}}
  \label{fig:field_idealized1}
 \end{figure*}

 \begin{figure*}
  \vspace{0.19cm}
  \begin{center}
    \includegraphics[width=0.4\textwidth]{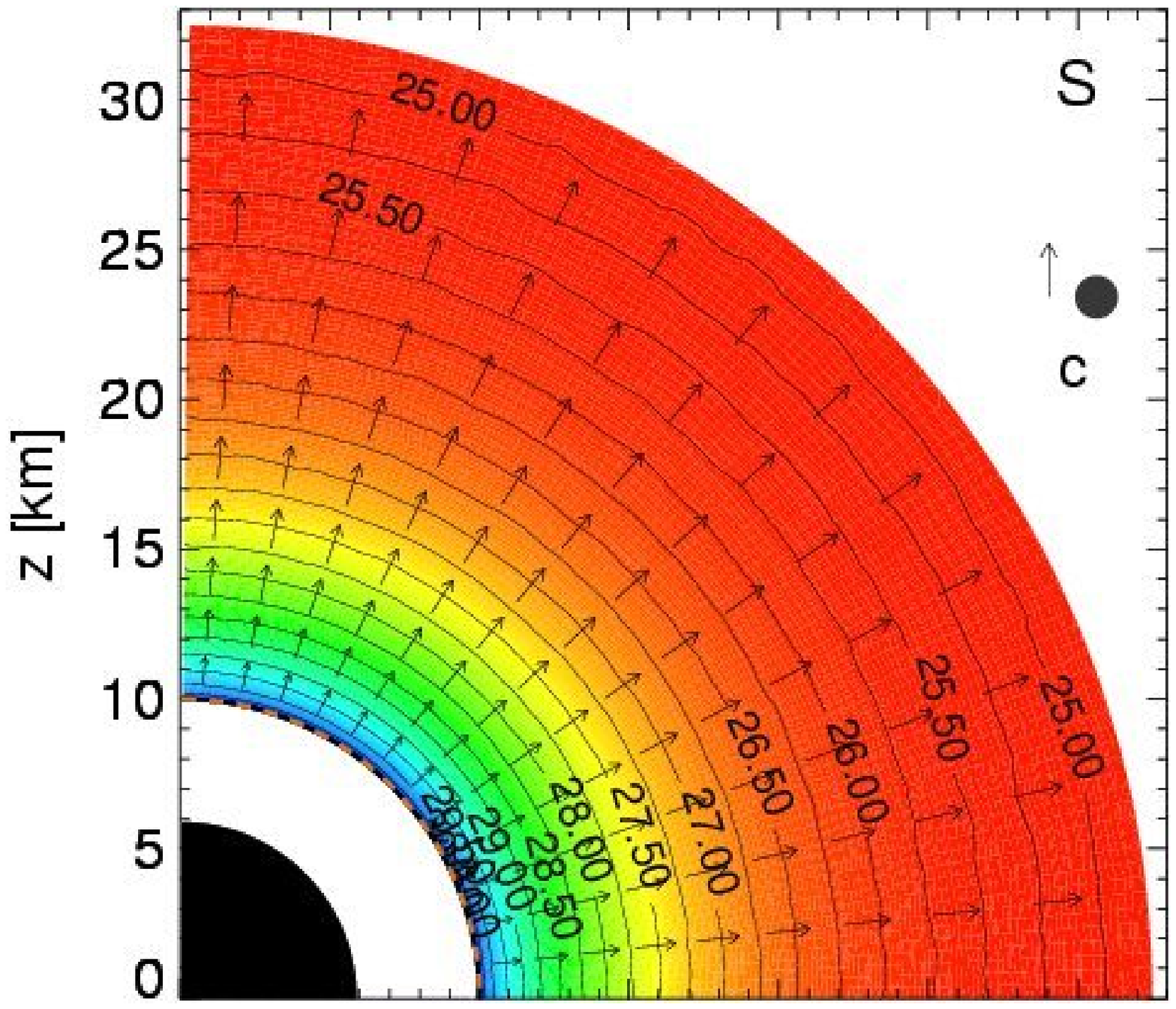}
   \hspace{1.75cm}
    \includegraphics[width=0.4\textwidth]{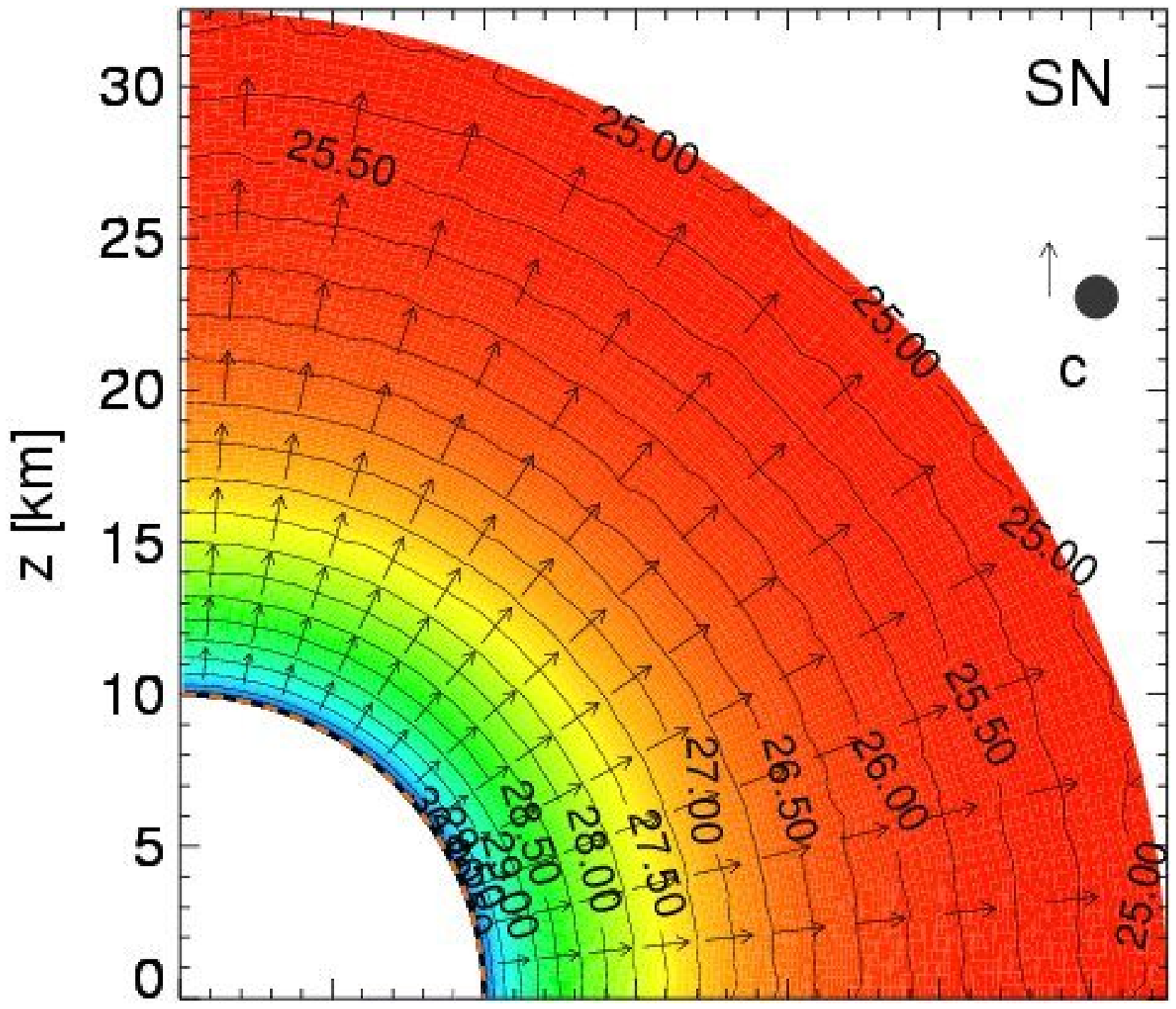}\\
   \vspace{0.6cm}
    \includegraphics[width=0.4\textwidth]{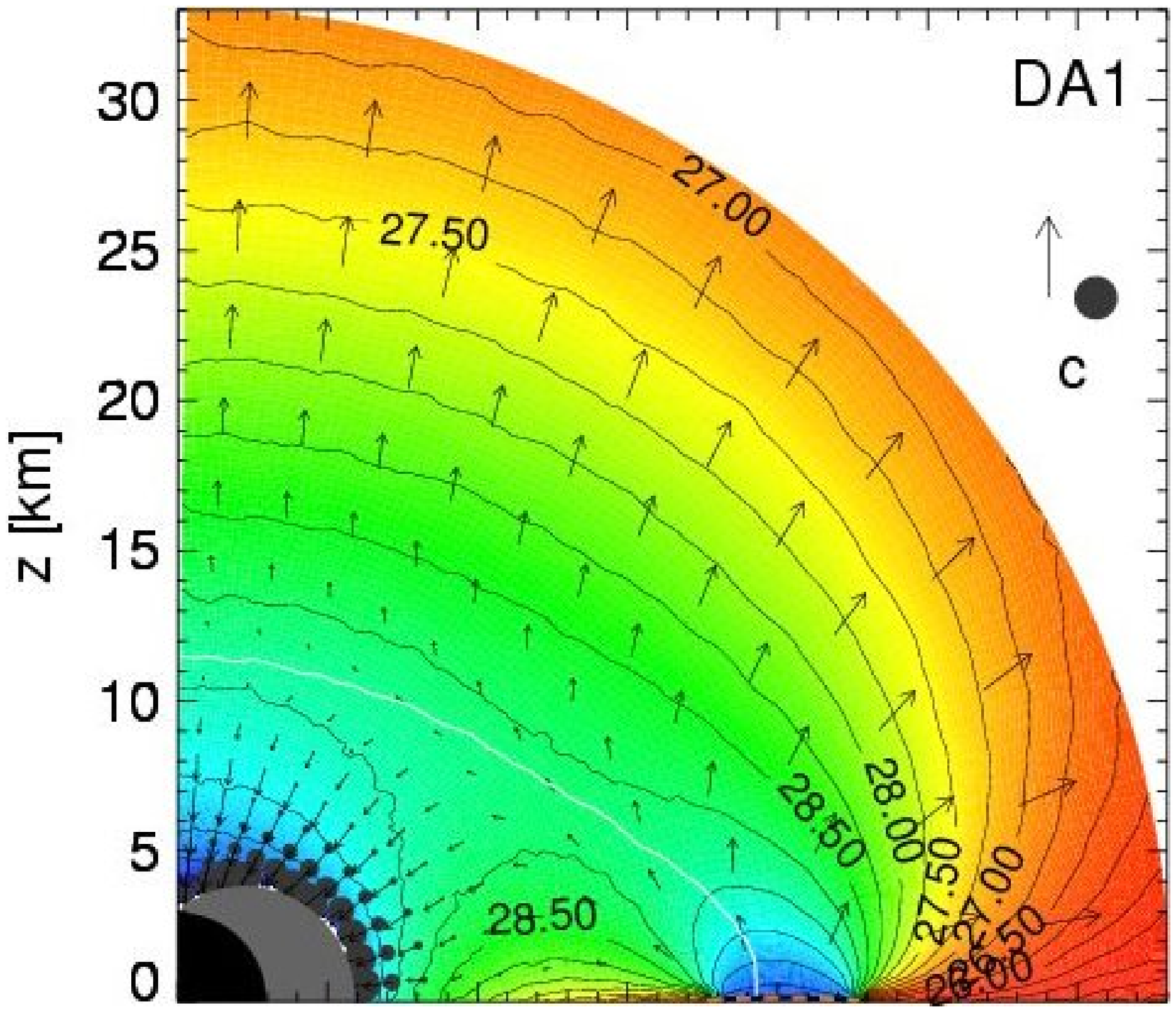}
   \hspace{1.75cm}
    \includegraphics[width=0.4\textwidth]{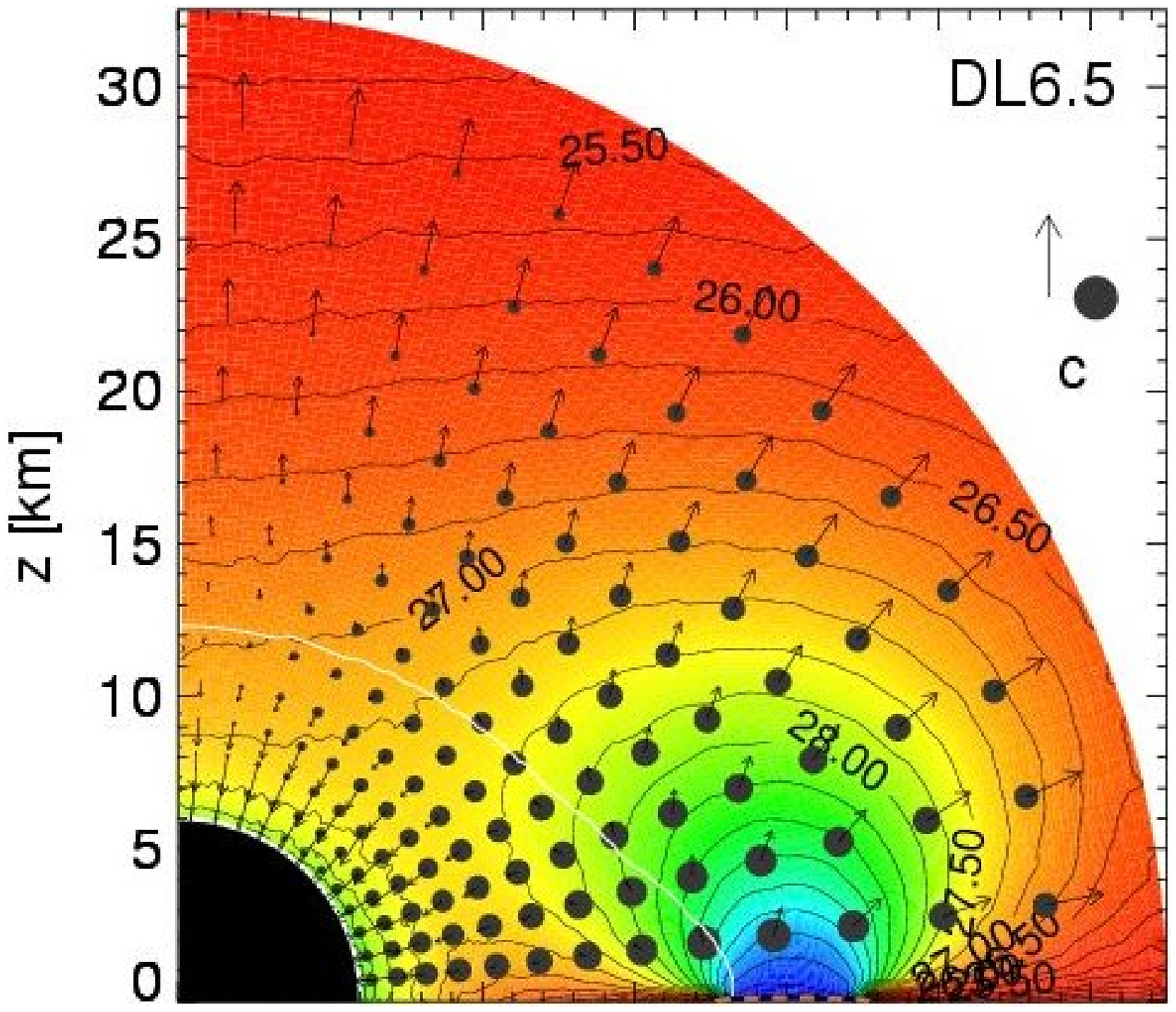}\\
   \vspace{0.6cm}
    \includegraphics[width=0.4\textwidth]{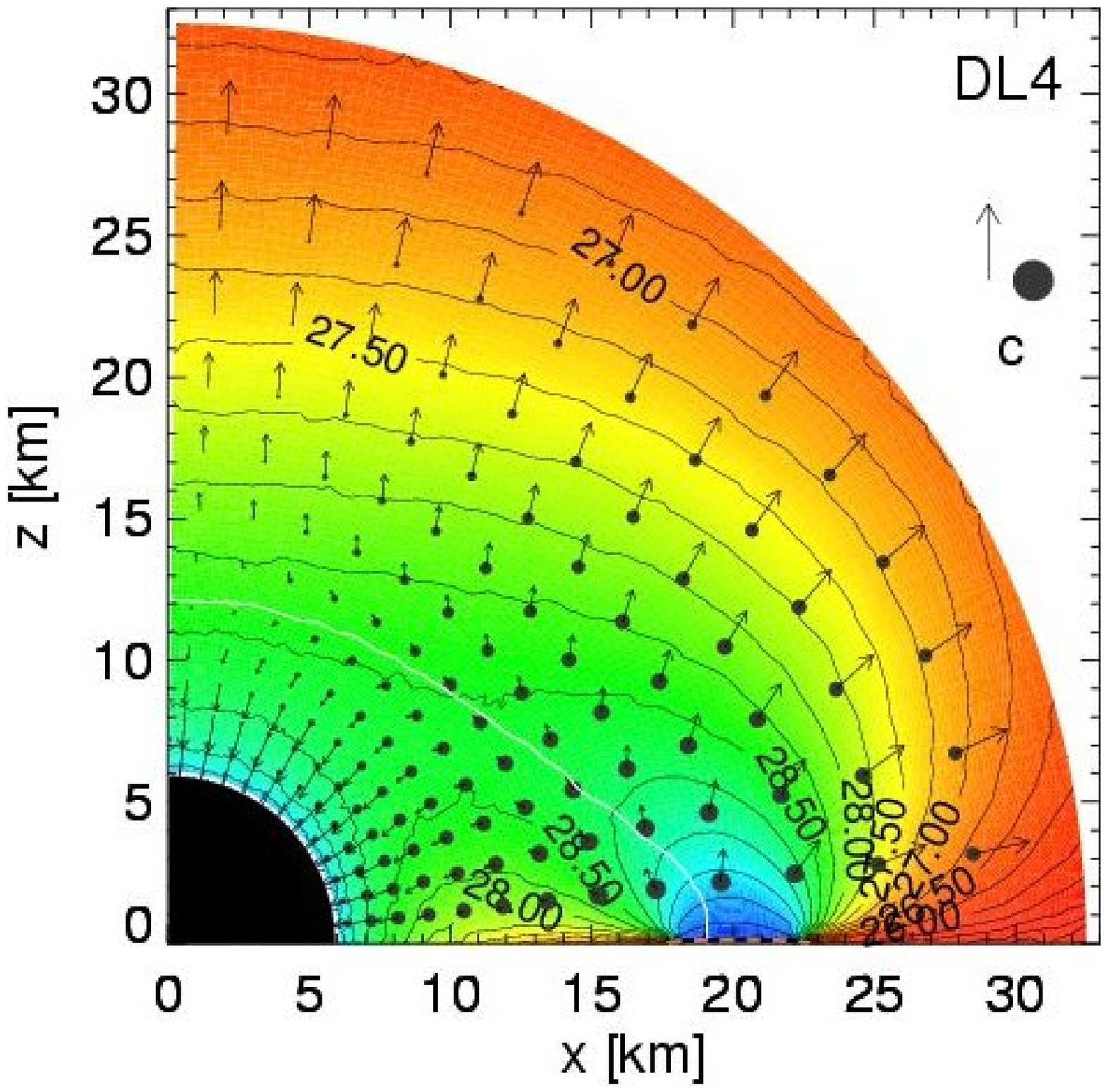}
   \hspace{1.75cm}
    \includegraphics[width=0.4\textwidth]{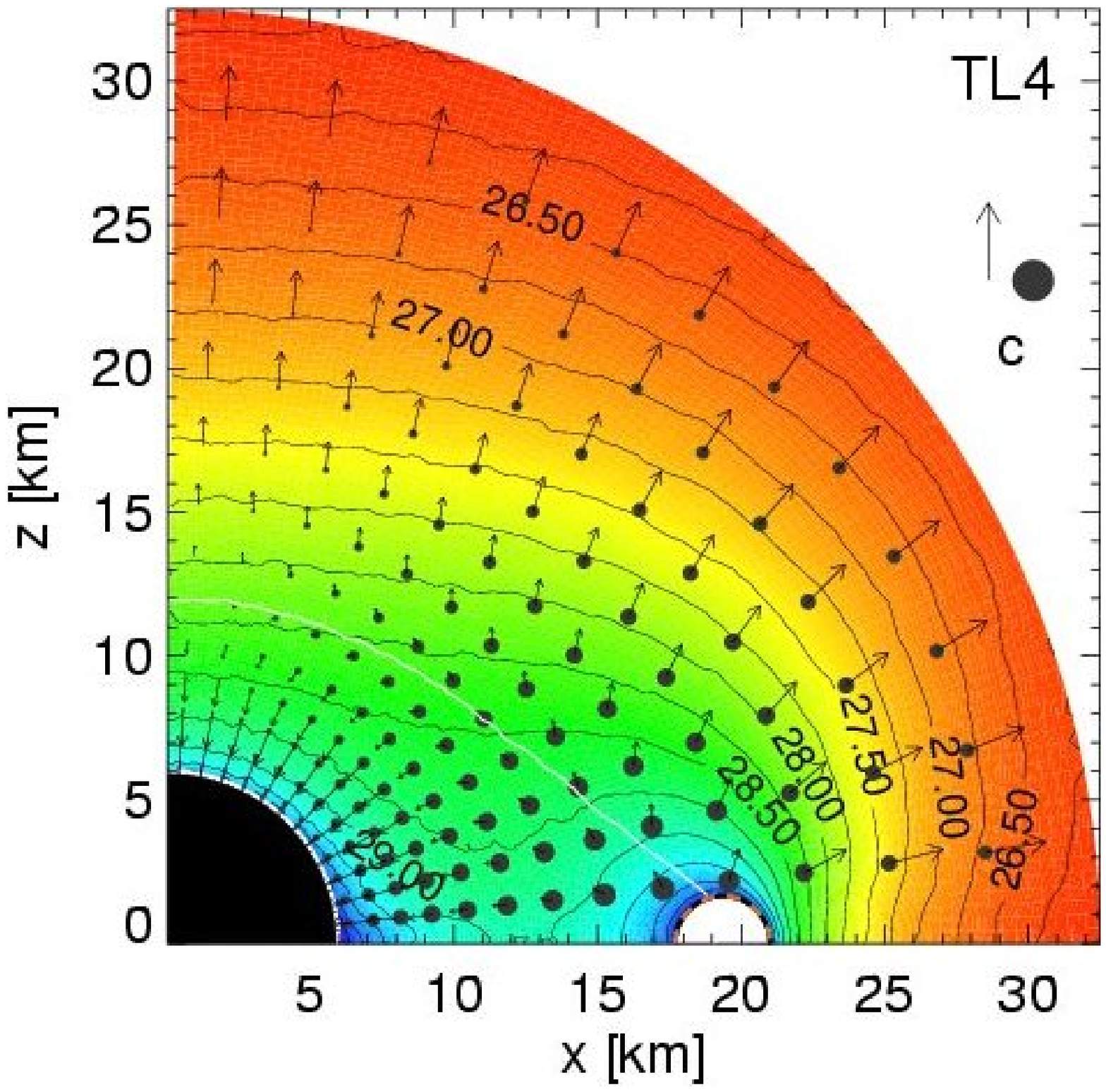}\\
   \vspace{0.3cm}
    \includegraphics{6293fg2e.eps}
  \end{center}
  \caption{Same as Fig.\,\ref{fig:field_idealized1}, but for a different set of models with
           idealized neutrinosphere geometries (see Table \ref{tab:idealized}). Note that model DA1
           contains a central rotating BH (with $a=1$) and therefore possesses an ergosphere outside
           of the event horizon. This ergosphere is marked by the gray area.
           {\em See the electronic edition for a color version of the figure.}}
  \label{fig:field_idealized2}
 \end{figure*}

 In the following we investigate several idealized BH-accretion disk systems. For each of these
 models the central black hole of mass $M$ and angular momentum $a$ is surrounded by a constructed
 neutrinosphere (see also Sect.\,\ref{sub:theory_neutrinosphere}), which is characterized by the
 geometry (thin disk, torus, or sphere), the position (radii), the rotation, and the temperature
 (see Table \ref{tab:idealized}). The rotation of the neutrinosphere is given by the uniform
 Lagrangian angular momentum $l = - u_{\phi} / u_{t}$ \citep{1978.Abramowicz}. This is the GR
 generalization of the specific angular momentum $l=\vec x\times\vec v$, where $\vec x$ is the
 position and $\vec v$ the velocity. In our models the prescribed geometry and position of the
 neutrinosphere limit physically reasonable values for the Lagrangian angular momentum $l$ to
 several times the mass of the BH (cgs units are obtained by applying the conversion factor $\left(
 M_{\rm GR}/M_{\odot}\right)\cdot4.42\cdot10^{15}\,{\rm cm}^2\,{\rm s}^{-1}$), and values of $l
 \approx 4\,M$ approximately describe Keplerian rotation. The neutrinosphere is assumed to be
 isothermal in the comoving frame, i.e. it has a uniform comoving frame temperature $T_{\rm C}$. In
 each idealized model the four parameters of the neutrinosphere presented here are also used for the
 antineutrinosphere.

 The idealized models are constructed for comparison with two reference models, which are the disk
 models D and REF (see Table \ref{tab:idealized}). Both models have the same temperature $T_{\rm
 C}$, and the disk is assumed to be non-rotating ($l=0$) for computing the \nnb-annihilation
 effects, but the masses $M$ of their non-rotating black holes and their outer disk radii differ.
 For each reference model there is a series of models (models DN to TL4N for model D and models A2B
 to L5 for model REF, see Table \ref{tab:idealized}), which are obtained by changing the properties
 of the corresponding reference model. We explore the influence of the geometry (toroidal models T,
 A2B, AB, A.5B, and TEMP, and spherical models S, and SM3), of GR effects by studying `Newtonian'
 models (disk model DN, torus model TN, and sphere model SN), of the inner disk radius (models
 RI4.5, and RI3), of the black hole angular momentum (models DA1, A.5, and A1), and of the disk
 rotation (models DL6.5, DL4, TL4, DL4N, TL4N, L2.5, L4, and L5).

 For most of the idealized models the effects due to the rotation of the neutrino emitting surface
 are ignored. Such models correspond to the cases which have $l=0$. Although in astrophysical
 systems disks and tori always rotate, we nevertheless consider idealized models with no
 neutrinosphere rotation here, because we want to selectively study the effects of disk or torus
 rotation on the neutrino-antineutrino annihilation rate. Later (see Sects.\,
 \ref{subsub:results_BH_rotation} and \ref{subsub:results_disk_rotation}), we will demonstrate that
 the rotation of the neutrino emitting source has a significant effect only on the spatial
 distribution of the annihilation rate but almost no influence on volume integrated results.

 The annihilation rate 4-vector $Q^{\alpha}$ of the series of models D to TL4 is shown in Figs.\,
 \ref{fig:field_idealized1} and \ref{fig:field_idealized2}. Similar plots can also be found in Fig.
 \,5 of \cite{2003.Miller}, and there especially the upper left panel ($a=0.0$, `MDR') nearly looks
 like our plot for model DL6.5 (see Fig.\,\ref{fig:field_idealized2}), where the accretion disk
 rotates with the super-Keplerian value $l=6.5$. The agreement is even somewhat better in case of
 model L5 (not visualized in our work), because in contrast to model DL6.5 it has the same inner
 (6\,$M$) and outer (10\,$M$) disk radii as the mentioned model of \cite{2003.Miller}. However, we
 were not able to reproduce the models of \cite{2003.Miller} exactly, because of the lack of
 information about the precise parameters of their models, particularly the rotation law. We also
 point out that in the pictorial representation of Fig.5 of \cite{2003.Miller}, the local value of
 the azimuthal integral of the `MDR' is provided, while our figures show the local value of $Q^t$.
 If instead of $Q^t$ we also compute the azimuthal integral of $Q^t$, the contours of, \eg model
 DL6.5 become vertical in the vicinity of the rotation axis just like in \cite{2003.Miller}.

 \begin{table}
  \tabcolsep=3.7mm
  \caption{Comparison of idealized models: For each pair of models $x_{\nu \bar\nu} \equiv \dot{E}_
           {\nu \bar\nu} \left({\rm M1}\right) / \dot{E}_{\nu \bar\nu} \left({\rm M2}\right)$
           denotes the ratio of the total annihilation rates in models M1 and M2. The superscript
           `tot' indicates that this ratio is obtained by integrating the local energy deposition
           rates over the whole grid, while for the superscript `up' the integration is restricted
           to the `up' region, where the energy released by \nnb-annihilation is not trapped by the
           BH and may eventually reach a distant observer. The ratio is computed in the local frame
           or in the frame of an infinitely distant observer. The latter case is indicated by the
           additional superscript `$\infty$'.
  \label{tab:comparison_idealized}}
  \begin{center}
   \begin{tabular}{cc|cccc}
    \hline\hline &&&&& \\*[-0.35cm]
     M1 & M2 &
     $x_{\nu\bar\nu}^{\rm tot}$ &
     $x_{\nu\bar\nu}^{\rm up}$ &
     $x_{\nu\bar\nu}^{{\rm tot},\infty}$ &
     $x_{\nu\bar\nu}^{{\rm up},\infty}$ \\*[0.05cm]
    \hline
     D     & DN  & 3.6  & 2.3  & 2.4  & 1.9  \\
     T     & TN  & 4.1  & 1.4  & 2.8  & 1.2  \\
     S     & SN  & 1.8  & 1.8  & 1.3  & 1.3  \\
     T     & D   & 1.2  & 0.48 & 1.2  & 0.48 \\
     S     & D   & 0.22 & 0.44 & 0.23 & 0.36 \\
     TN    & DN  & 1.1  & 0.79 & 1.1  & 0.79 \\
     SN    & DN  & 0.44 & 0.55 & 0.44 & 0.55 \\
     DA1   & D   & 1.0  & 1.0  & 1.1  & 1.0  \\
     DL6.5 & D   & 0.56 & 1.0  & 0.69 & 1.0  \\
    \hline
     A2B   & REF & 0.92 & 0.61 & 0.81 & 0.61 \\
     AB    & REF & 1.1  & 0.45 & 0.92 & 0.45 \\
     A.5B  & REF & 1.7  & 0.39 & 1.2  & 0.39 \\
     RI4.5 & REF & 1.5  & 1.3  & 1.3  & 1.2  \\
     RI3   & REF & 2.3  & 1.3  & 1.8  & 1.3  \\
     TEMP  & AB  & 5.3  & 3.0  & 4.7  & 2.9  \\ \hline
   \end{tabular}
  \end{center}
 \end{table}

 \begin{figure*}
  \begin{tabular}{cc}
   \includegraphics[width=0.48\textwidth]{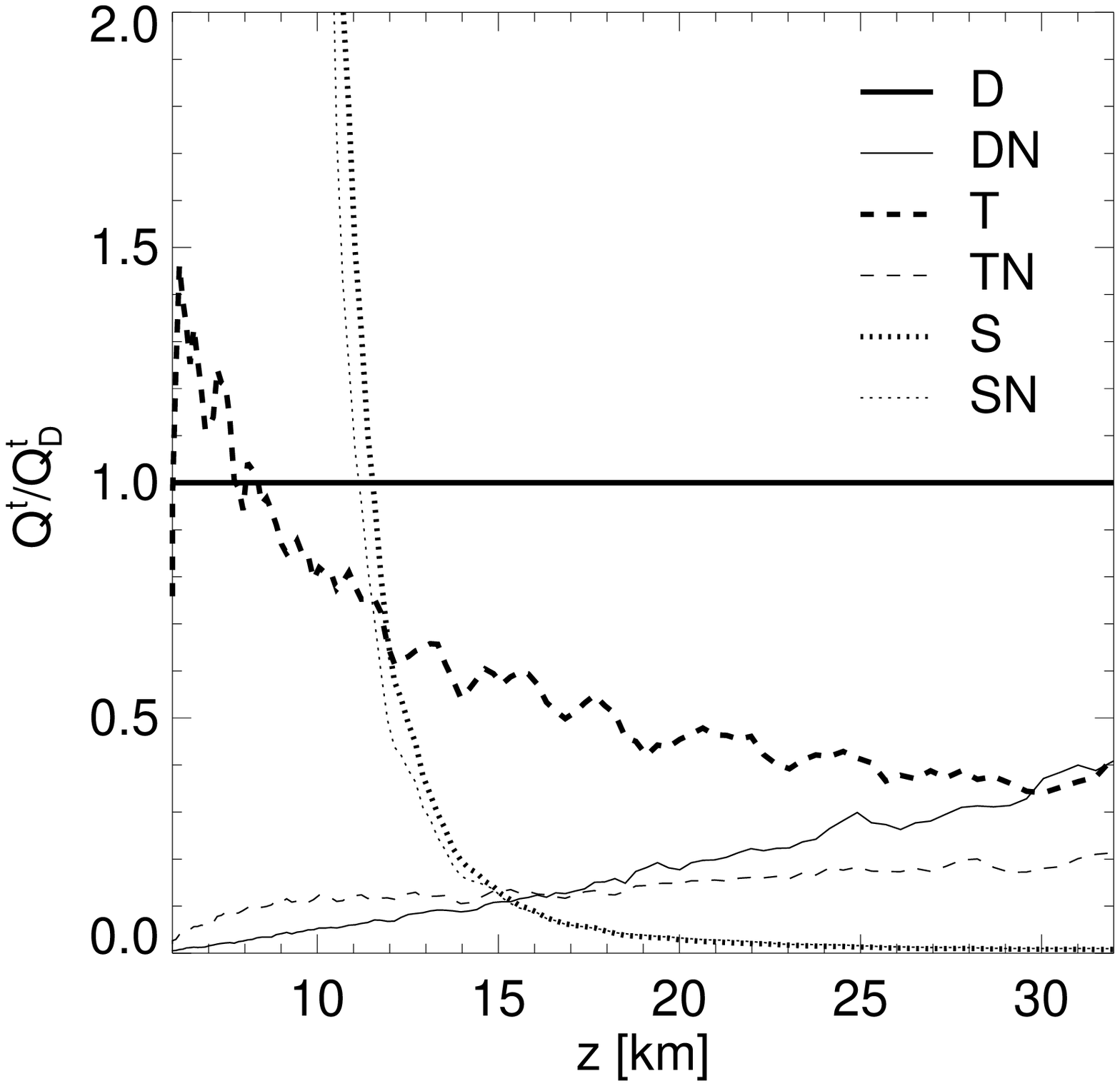} &
   \includegraphics[width=0.48\textwidth]{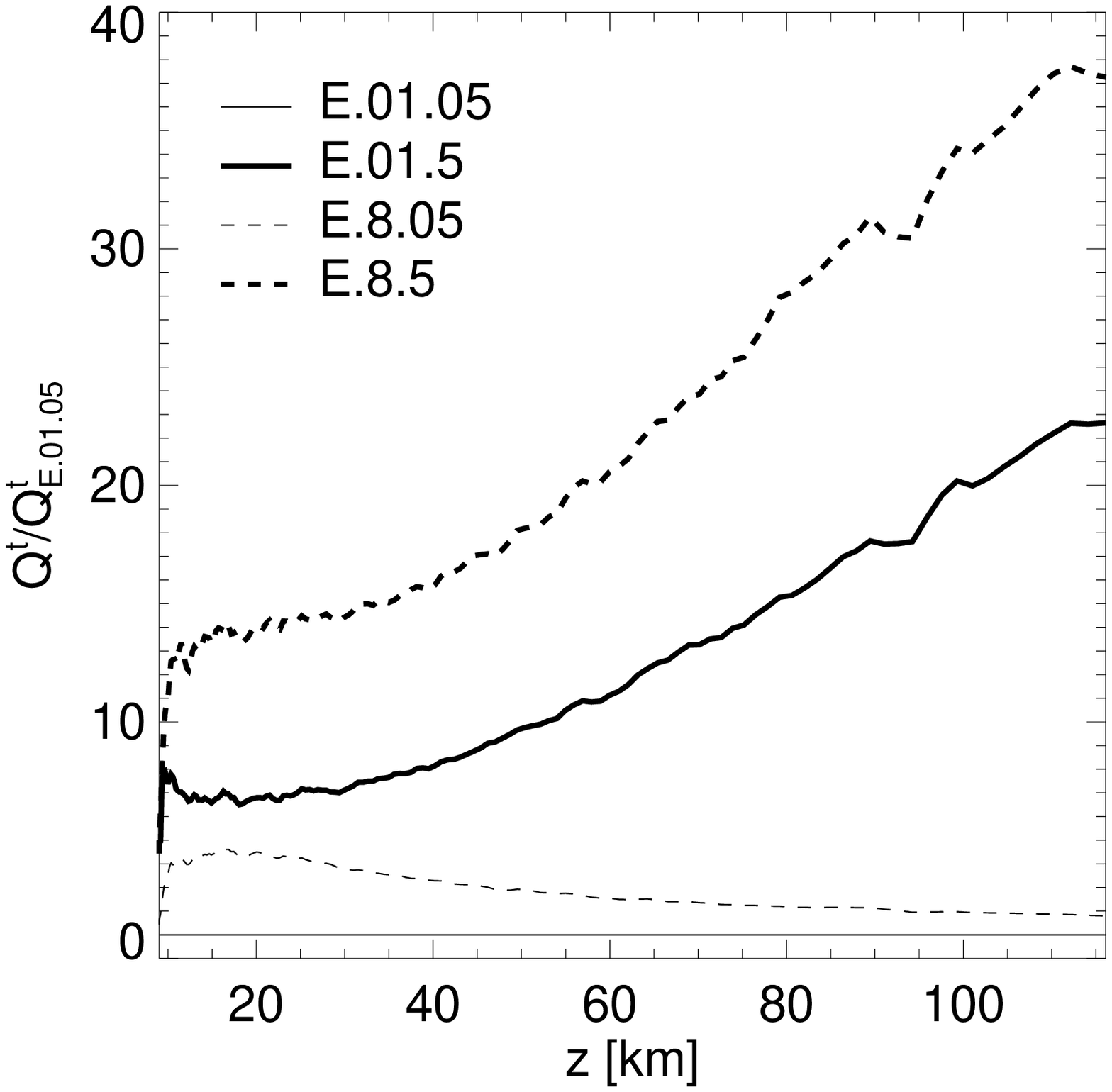}
  \end{tabular}
  \caption{Energy component of the annihilation rate 4-vector along the z-axis for models with
           idealized neutrinosphere geometries (left panel; see Table \ref{tab:idealized}), and for
           models where the neutrinospheres are calculated from equilibrium torus models (right
           panel, see Table \ref{tab:equilibrium}). The values are normalized to the energy
           component of the annihilation rate 4-vector along the z-axis of models D and E.01.05,
           respectively.}
  \label{fig:z-axis}
 \end{figure*}

 \begin{figure*}
  \begin{tabular}{cc}
   \includegraphics[width=0.48\textwidth]{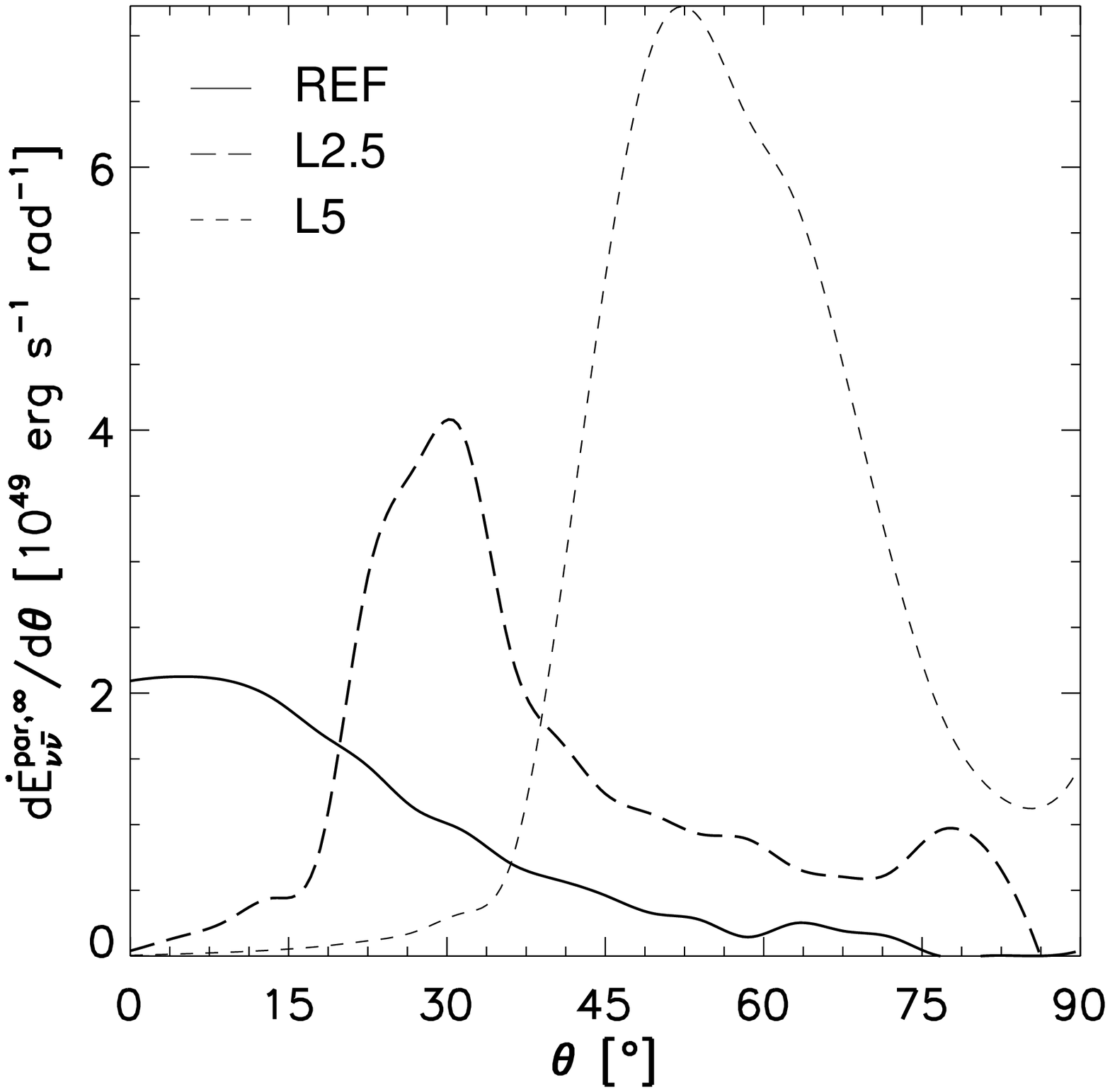} &
   \includegraphics[width=0.48\textwidth]{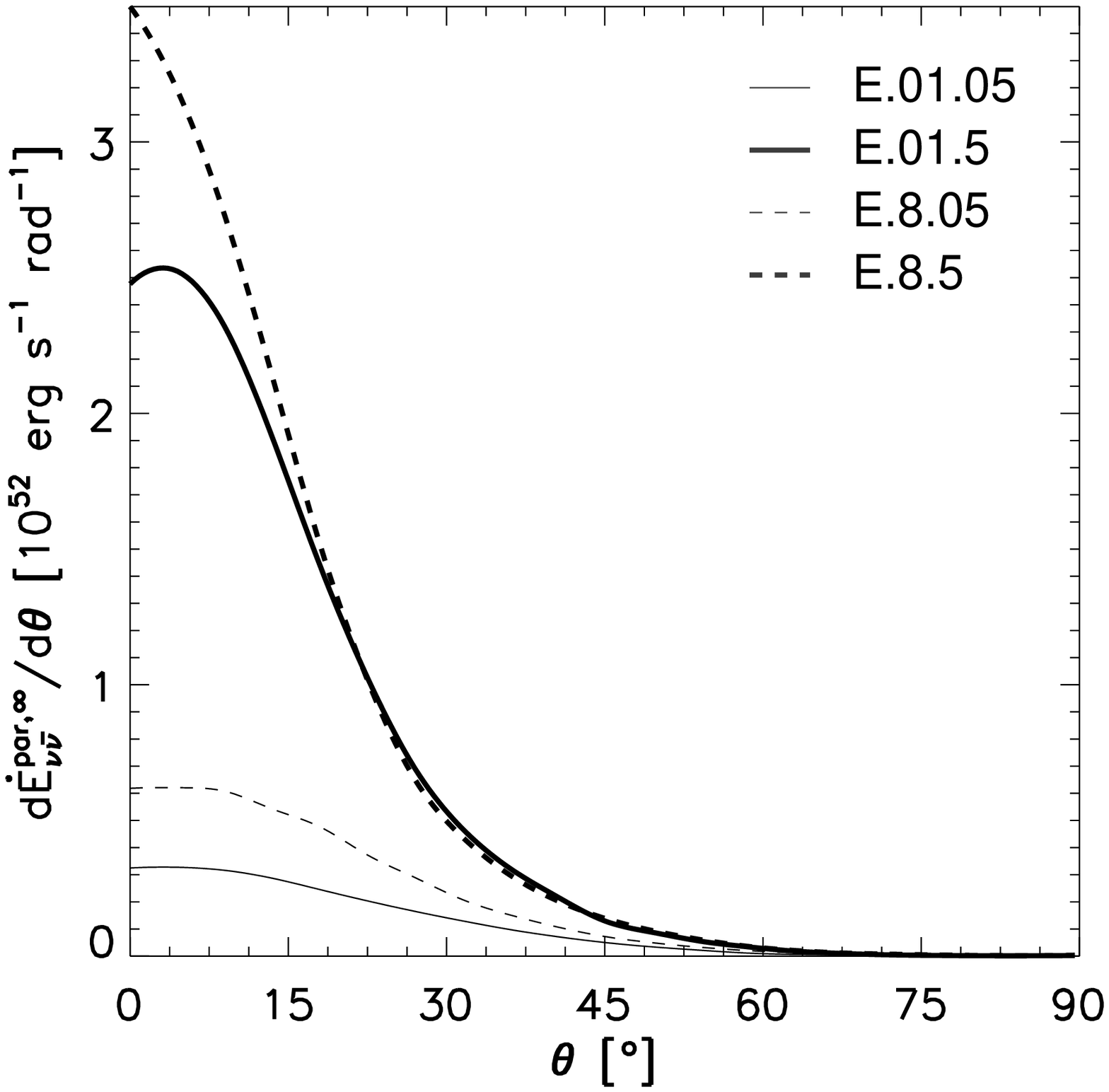}
  \end{tabular}
  \caption{Energy deposition rate $\dot{E}_{\nu\bar{\nu}}^{{\rm par},\infty}$ on a sphere of radius
           $200\,M$ per unit of Boyer-Lindquist polar angle $\theta$, obtained by taking into
           account the energy transport to this radius as described by the annihilation rate
           4-vector field $Q^{\alpha}$. The left panel shows the results for idealized disk models
           with different values of the Lagrangian angular momentum $l$, and the right panel the
           results for the first four equilibrium torus models of Table
           \ref{tab:equilibrium}.}
  \label{fig:angular}
 \end{figure*}

\subsubsection{Influence of general relativity}

 In order to study the influence of GR effects, we take the thin disk model D, the torus model T,
 and the sphere model S, and compare these models with their corresponding Newtonian cases DN, TN,
 and SN (see Table \ref{tab:idealized}). The Newtonian models have the same neutrinosphere as the
 corresponding GR models, however, we assume $M=0$, i.e. we neglect the influence of gravity on the
 neutrino propagation and transformation between observers. The computational grid for a Newtonian
 model is identical to that of the corresponding GR model, i.e. in particular, the inner radial grid
 boundary is located at the same radius in both models (just above the horizon of the GR model). All
 three Newtonian models DN, TN, and SN are assumed to have the same surface area and temperature,
 and therefore also have the same total (neutrino plus antineutrino) luminosity $L_{\nu}^{{\rm N}}$,
 i.e. $L_{\nu,{\rm DN}} = L_{\nu,{\rm TN}} = L_{\nu,{\rm SN}} = 3.9 \cdot 10^{51}\,$erg\,s$^{-1}$
 (see Table \ref{tab:idealized}). Although the GR models D, T, and S have the same metric ($M=2\,M_
 {\odot},a=0$), a local comoving observer does not measure the same surface areas, because the
 surfaces of these three models do not coincide such that they are differently affected by the
 curvature of spacetime. Therefore, despite having the same temperature ($T_{\rm C}=5\cdot10^
 {10}\,K$), models D, T, and S have a different local luminosity (see Eq.\,\ref{eq:Lnu}). Note that
 in the Newtonian models the energy emission rate from the neutrinosphere and the energy deposition
 rates do not depend on the observer, because there is no gravitational redshift effect.

 In the local frame GR effects increase the energy deposition rate $\dot{E}_{\nu \bar\nu}$ of models
 D, T, and S with respect to their Newtonian counterparts (see Table
 \ref{tab:comparison_idealized}). The increase is smaller when restricting the comparison to the
 `up' region (where the radial component of the momentum deposition vector is positive, $Q^r>0$).
 But even in the `worst' case (toroidal model T) GR effects enhance $\dot{E}_{\nu \bar\nu}^{\rm up}$
 by about 40\%. The optimal geometry to release energy in the surroundings of the BH is that of a
 thin disk, where the GR enhancement of the energy deposition rate in the `up' region is $x_{\nu\bar
 \nu}^{\rm up} = 2.3$ for a local observer and where the energy deposition rate in the `up' volume
 is larger than for toroidal and spherical geometries. These results also hold for a distant
 observer (Table \ref{tab:comparison_idealized}). The energy deposition by \nnb-annihilation is
 therefore enhanced in GR models compared to the Newtonian treatment, despite the fact that the
 luminosity $L^{\infty}_{\nu}$ for an observer at infinity is smaller (Table \ref{tab:idealized}).
 This in the first moment counterintuitive behaviour can be understood by the relativistic effects,
 which account for a more intense neutrino radiation field close to the emitting surface, i.e. the
 local luminosity, which is not reduced by gravitational redshift, is always higher or at least
 equal to the Newtonian counterpart of a GR model.

 The enhancement of the energy deposition rate due to GR effects depends strongly on the position,
 and becomes negligible at sufficiently large distances from the BH (Fig.\,\ref{fig:z-axis}, left
 panel), where relativistic effects are unimportant. In case of the disk models (D and DN), Fig.\,
 \ref{fig:field_idealized1} (upper panels) and Fig.\,\ref{fig:z-axis} (left panel) show that GR
 effects enhance the energy deposition rate near the symmetry axis ($\theta \lsim 30^o$) by a factor
 of $\gsim\,10$ up to $r \sim 10\,$km, and they still increase the energy release by a factor of
 $\sim 2$ at $r \sim 30\,$km.

\subsubsection{Influence of the neutrinosphere geometry}

 According to Table \ref{tab:comparison_idealized}, the energy deposition rate by \nnb-annihilation
 in the `up' region measured by an observer at infinity is largest for a neutrinosphere having the
 geometry of a thin disk (model D; $\dot{E}_{\nu\bar\nu}^{\rm up, \infty} = 2.3 \cdot 10^{48}\,$erg
 \,s$^{-1}$; Table \ref{tab:idealized}), while it is smallest for a neutrinosphere of spherical
 shape (model S; $\dot{E}_{\nu \bar\nu}^{\rm up, \infty} = 0.83 \cdot 10^{48}\,$erg\,s$^{-1}$;
 Table \ref{tab:idealized}). This finding holds for the corresponding Newtonian models (DN and SN),
 too. This is a remarkable result, as in the spherical models $Q^r>0$ everywhere on the
 computational grid (for models S and SN the \nnb-annihilation integrals in the `up' regions are
 equal to the corresponding integrals in the `tot' region, respectively; see Table
 \ref{tab:idealized}).

 The influence of the neutrinosphere geometry on the energy deposition rate can also be studied by
 considering toroidal models with different meridional cross sections. All our toroidal models have
 an elliptical meridional cross section characterized by its eccentricity (Table
 \ref{tab:idealized}). Model REF (an infinitely thin disk model with an eccentricity $e=1$) serves
 as the reference model, which is compared to models A2B ($e=\sqrt{3}/2$, oblate), AB ($e=0$,
 circular) and A.5B ($e=\sqrt{3}/2$, prolate), respectively. These four models have the same
 Newtonian luminosity $L_{\nu}^{{\rm N}}$, and the center of their meridional cross sections or the
 mid-points of their disks are all located at $r_{\rm c} \approx 18\,$km in the equatorial plane.
 Compared to the thin disk model REF, the other torus models are all less efficient in depositing
 energy in the `up' region (Table \ref{tab:comparison_idealized}). By inspection of Table
 \ref{tab:idealized} one recognizes that the energy deposition rate in the `up' region measured
 locally and by a distant observer $\dot{E}_{\nu\bar\nu}^{{\rm up}, \infty}$ decreases when the
 meridional cross section becomes more prolate. It is smallest for model A.5B, which is the most
 prolate model. At the same time this model is the most efficient one in the total energy release
 ($x_{\nu \bar{\nu},{\rm A.5B}}^{{\rm tot}, \infty} = 1.2 > 1$). This at first glance surprising
 behavior can be explained by the fact that a thin disk emits a large fraction of the neutrinos in
 directions parallel to the symmetry axis, and hence just above and below the thin disk.
 Consequently, \nnb-encounters are less frequent close to the BH than in a source of toroidal shape.
 Although infinitely thin disks are a mathematical idealization more than a natural case, our
 results nevertheless imply that a distant observer receives more energy from systems which have a
 more oblate structure of a toroidal accretion disk.

 Another geometrical aspect of relevance is the size of the innermost radius of the disk ($r_{\rm
 in}$). To study the dependence of the energy deposition rate on this parameter, we consider the
 thin disk models RI4.5 ($r_{\rm in} = 4.5M$) and RI3 ($r_{\rm in} = 3M$), which both have an
 innermost radius smaller than that of model REF ($r_{\rm in} = 6M$), as can be seen from Table
 \ref{tab:idealized}, but have the same Newtonian luminosity $L_{\nu}^{{\rm N}}$ and Newtonian
 surface area. Independent of the region of integration (`up' or `total') and of the observer (local
 or distant), we find that the smaller is $r_{\rm in}$, the larger is the energy released by
 \nnb-annihilation (Table \ref{tab:comparison_idealized}).

 Finally, we consider model TEMP (Table \ref{tab:idealized}) to further investigate the influence of
 the neutrinosphere geometry on the energy deposition rate. This model is identical to model AB,
 except that in model TEMP only the inner half of the torus (up to a distance of $8M$ from the
 symmetry axis) is assumed to emit neutrinos. Thus, in order to have the same Newtonian luminosity
 $L^{\rm N}_\nu$ in both models, the neutrinosphere temperature has to be increased from $T_C = 5
 \cdot 10^{10}\,$K to $T_C = 6.1\cdot 10^{10}\,$K, because the emitting surface is smaller. The more
 concentrated neutrino radiation field therefore leads to a strong overall increase of the energy
 deposition rate ($x_{\nu\bar\nu} \gsim 3$; Table \ref{tab:comparison_idealized}), in analogy to the
 models of the previous paragraph, where the innermost disk radius was reduced.

 \begin{table*}
  \tabcolsep=1.2mm
  \caption{Properties of equilibrium torus models, where the location of the $\nu_{\rm e}$-sphere
           differs from that of the $\bar{\nu}_{\rm e}$-sphere. Besides the quantities already
           defined in Table \ref{tab:idealized}, the models here are characterized by the mass $m_{
           \rm tor}$ of the accretion torus, and the mass $M_{\rm tor}$ and angular momentum $a_{\rm
           tor}$ of the BH used to calculate the structure of the accretion torus. In most models
           the values of the last two quantities are equal to the values of the mass $M$ and angular
           momentum $a$ of the BH that we use to determine the neutrino geodesics. However, in the
           two models E.01.1d and E1.1d we have chosen $a_{\rm tor}\neq a$, in order to discriminate
           between the influence of the black hole rotation on the accretion torus and on the
           neutrino trajectories (see Sect.\,\ref{sub:results_equilibrium}). For the Newtonian model
           E.01.1N the table shows $M\neq M_{\rm tor}$ (and also $a\neq a_{\rm tor}$), because the
           neutrino trajectories are evaluated without GR effects. All models in this table have the
           same photon entropy $s_{\gamma}=1$, the same inner equatorial torus radius $r_{\rm in}=
           4.1\,M$, and the same electron fraction $Y_{\rm e}=0.1$ (see
           Sect.~\ref{sub:theory_neutrinosphere}). Moreover, the effects of the torus angular
           momentum $l$ are included in the evaluation of the \nnb-annihilation. Note that $l$ is
           derived from the set of six quantities $(M,a,m_\mathrm{tor},s_\gamma,r_\mathrm{in},
           Y_{\rm e})$ that fully specify a torus in rotational equilibrium.
  \label{tab:equilibrium}}
  \begin{center}
   \begin{tabular}{c|cc|ccc|cc|cccccc}
    \hline\hline &&&&&&&&&&&&& \\*[-0.35cm]
     model & $M$ & $a$ & $M_{{\rm tor}}$ &
     $a_{{\rm tor}}$ & $m_{{\rm tor}}$ & $l$ & $L^{\infty}_{\nu}$ &
     $\dot{E}_{\nu\bar{\nu}}^{{\rm tot}}$ &
     $\dot{E}_{\nu\bar{\nu}}^{{\rm up}}$ &
     $\dot{E}_{\nu\bar{\nu}}^{{\rm tot},\infty}$ &
     $\dot{E}_{\nu\bar{\nu}}^{{\rm up},\infty}$ &
     $q_{\nu\bar{\nu}}^{{\rm tot},\infty}$ &
     $q_{\nu\bar{\nu}}^{{\rm up},\infty}$ \\
     name & $M_{\odot}$ && $M_{\odot}$ & $M$ & $M_{\odot}$ & $M_{\rm GR}$ &
     $10^{54}$\ergsec &
     $10^{52}$\ergsec &
     $10^{52}$\ergsec &
     $10^{52}$\ergsec &
     $10^{52}$\ergsec &
     $10^{-2}$ & $10^{-2}$ \\*[0.05cm]
    \hline
     E.01.05 & 3 & 0.01 & 3 & 0.01 & 0.05 & 3.95 & 0.32 &  0.83 & 0.40 & 0.59 & 0.36 & 1.9 & 1.1 \\
     E.01.5  & 3 & 0.01 & 3 & 0.01 & 0.5  & 3.98 & 0.99 &  6.4  & 3.0  & 4.6  & 2.7  & 4.7 & 2.7 \\
     E.8.05  & 3 & 0.8  & 3 & 0.8  & 0.05 & 3.48 & 0.44 &  2.9  & 1.0  & 1.9  & 0.93 & 4.3 & 2.1 \\
     E.8.5   & 3 & 0.8  & 3 & 0.8  & 0.5  & 3.53 & 1.1  & 10    & 4.0  & 7.4  & 3.7  & 6.6 & 3.3 \\
    \hline
     E.01.1  & 3 & 0.01 & 3 & 0.01 & 0.1  & 3.96 & 0.51 &  2.1  & 1.0  & 1.5  & 0.92 & 2.9 & 1.8 \\
     E.01.1d & 3 & 0.01 & 3 & 1    & 0.1  & 3.96 & 0.62 &  4.5  & 1.7  & 2.9  & 1.5  & 4.6 & 2.4 \\
     E1.1d   & 3 & 1    & 3 & 0.01 & 0.1  & 3.96 & 0.51 &  2.0  & 1.0  & 1.6  & 0.94 & 3.1 & 1.8 \\
     E1.1    & 3 & 1    & 3 & 1    & 0.1  & 3.48 & 0.62 &  4.9  & 1.7  & 3.3  & 1.5  & 5.3 & 2.4 \\
    \hline
     E.01.25 & 3 & 0.01 & 3 & 0.01 & 0.25 & 3.98 & 0.79 &  4.6  & 2.1  & 3.2  & 1.9  & 4.1 & 2.4 \\
     E.01.1N & 0 & 0    & 3 & 0.01 & 0.1  & 3.96 & 0.57 &  0.90 & 0.76 & 0.90 & 0.76 & 1.6 & 1.3 \\
    \hline
   \end{tabular}
  \end{center}
 \end{table*}

 \begin{figure*}
  \vspace{0.19cm}
  \begin{center}
    \includegraphics[width=0.4\textwidth]{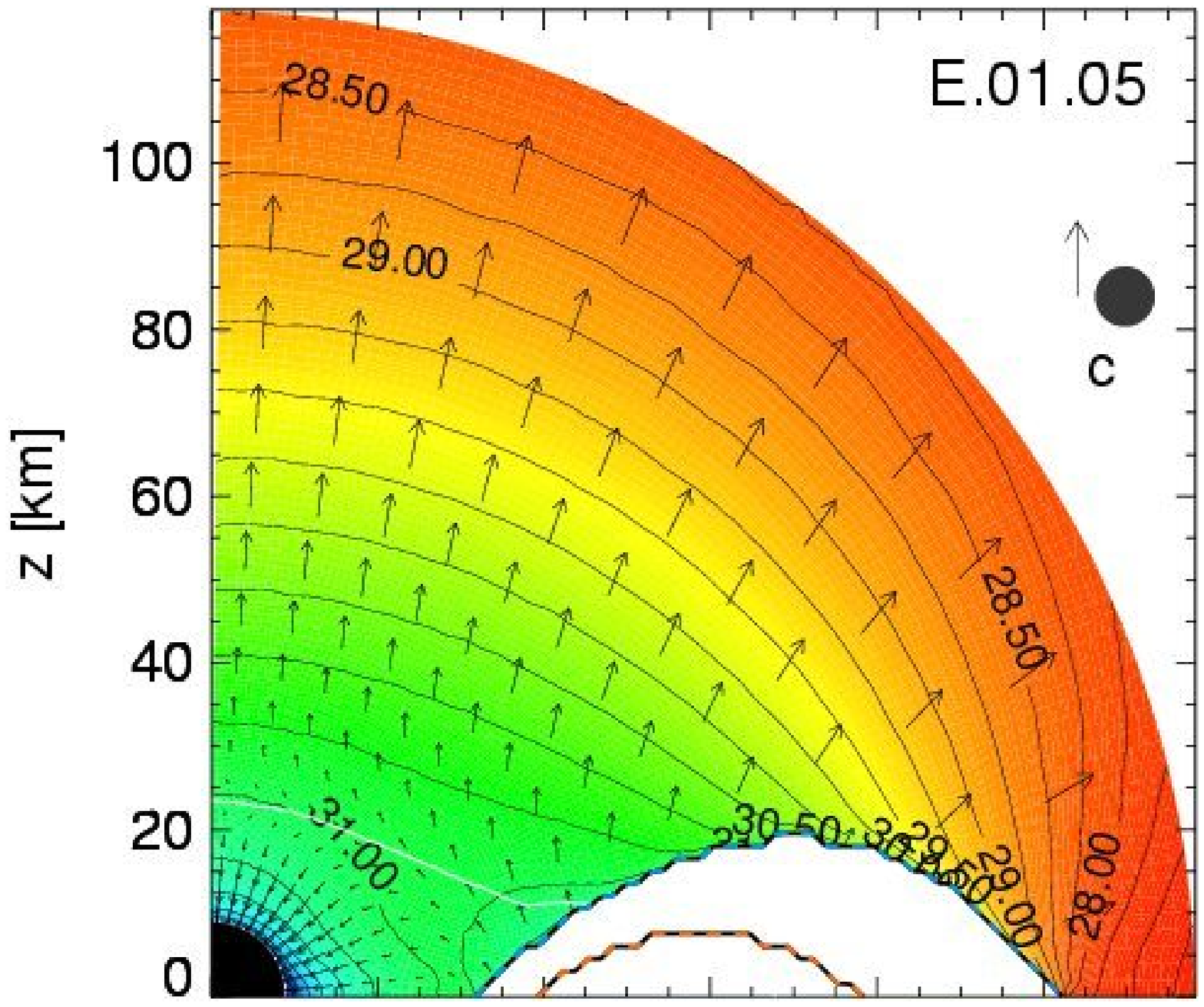}
   \hspace{1.75cm}
    \includegraphics[width=0.4\textwidth]{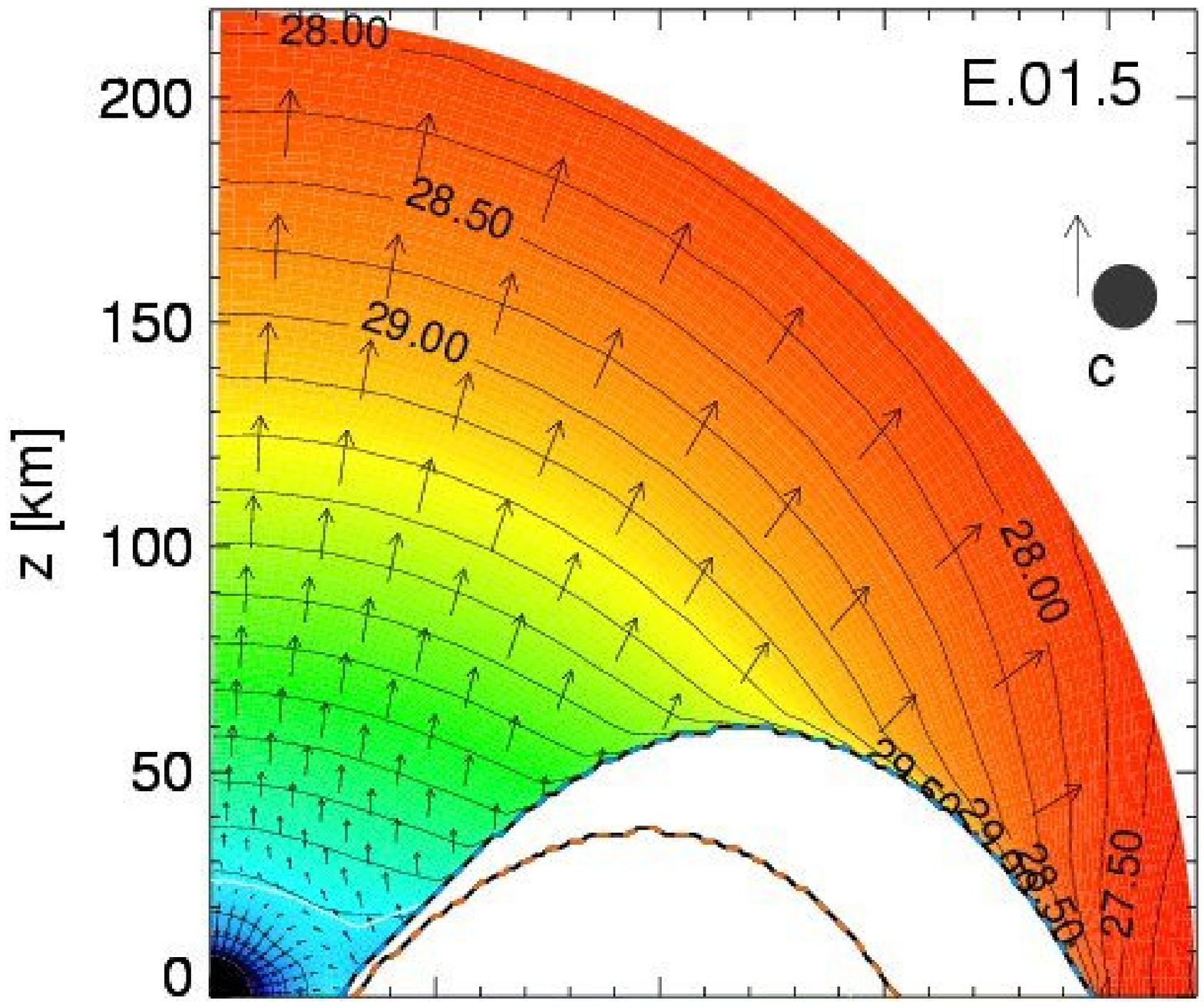}\\
   \vspace{0.6cm}
    \includegraphics[width=0.4\textwidth]{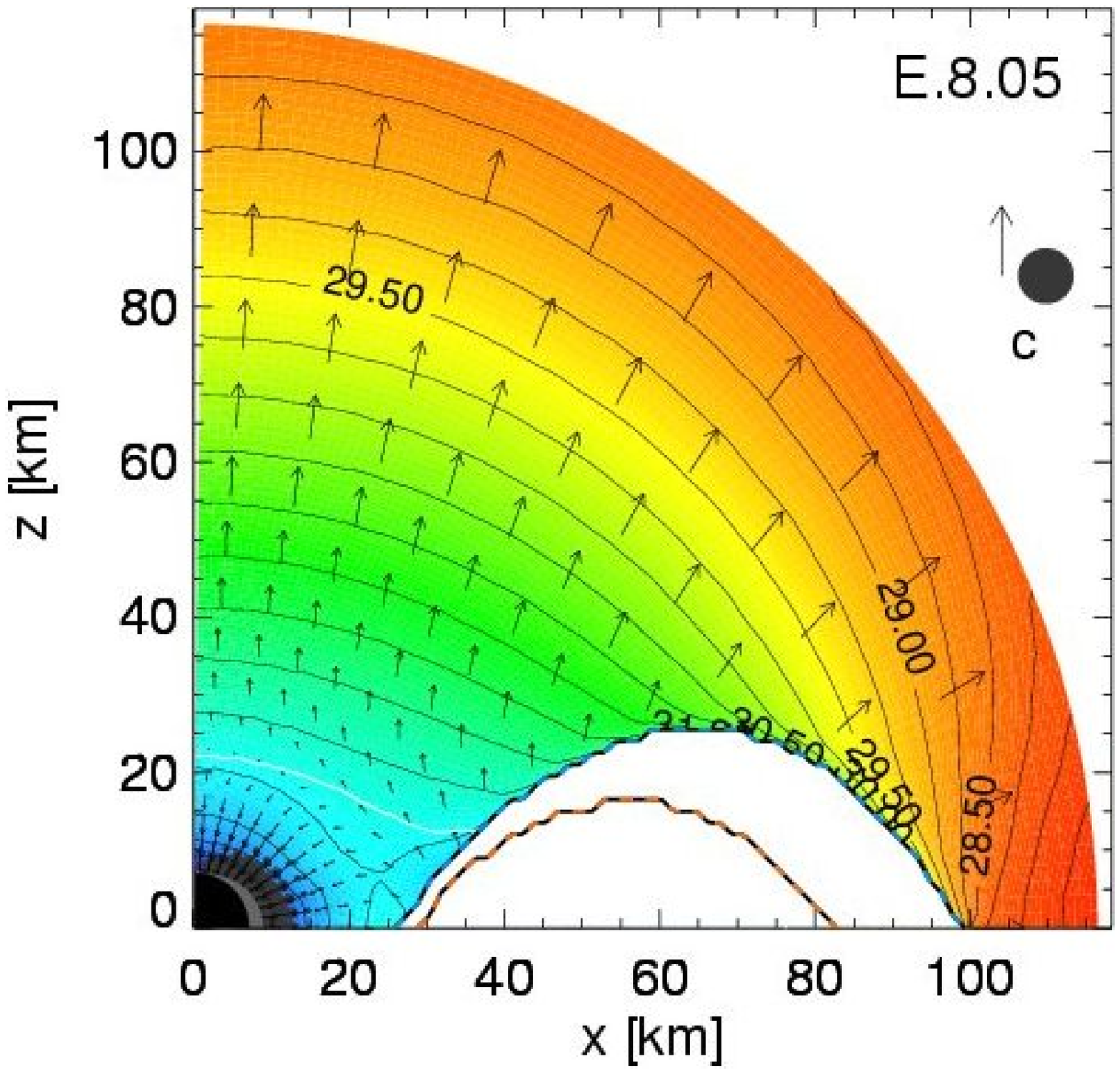}
   \hspace{1.75cm}
    \includegraphics[width=0.4\textwidth]{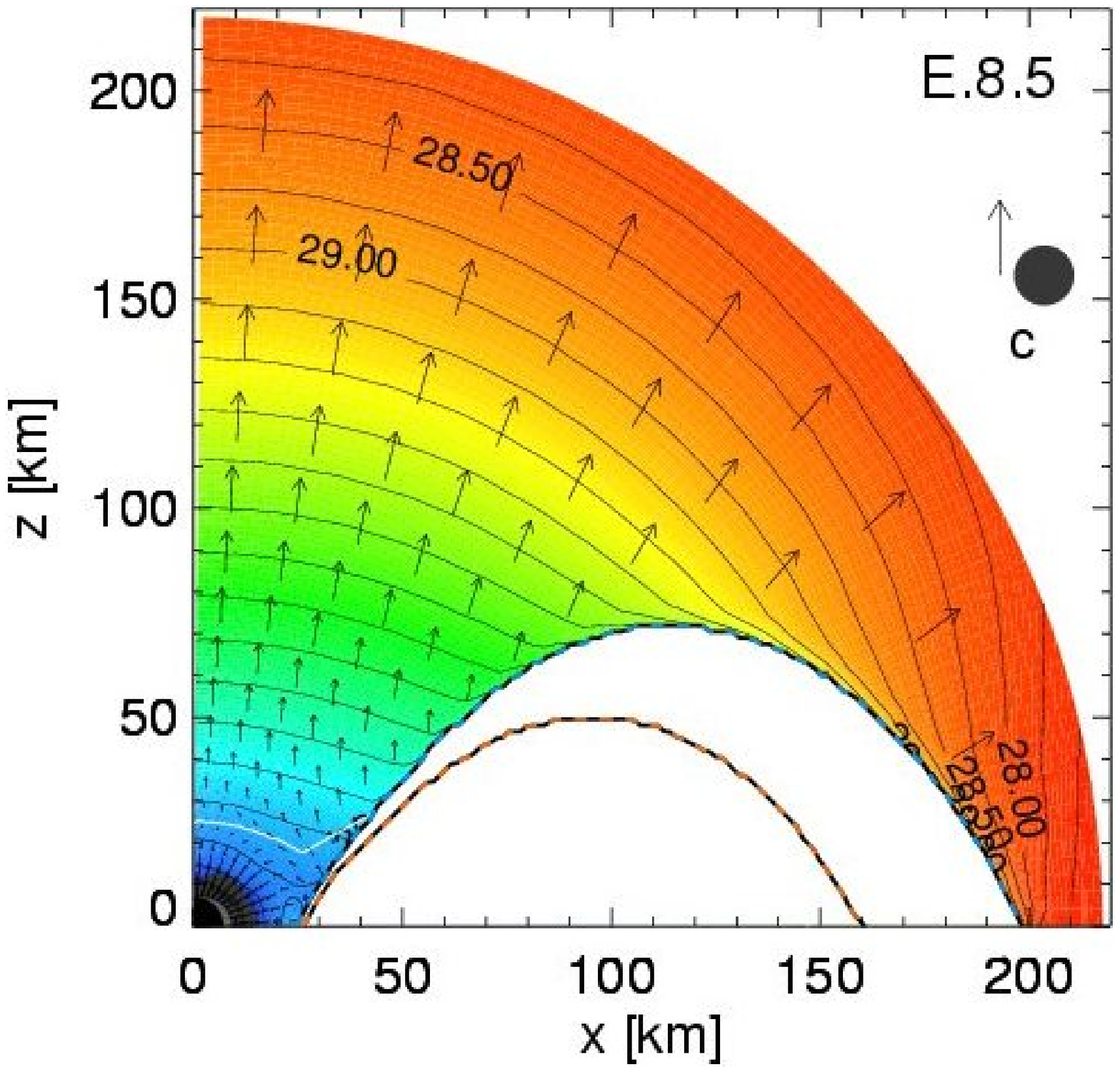}\\
   \vspace{0.3cm}
    \includegraphics{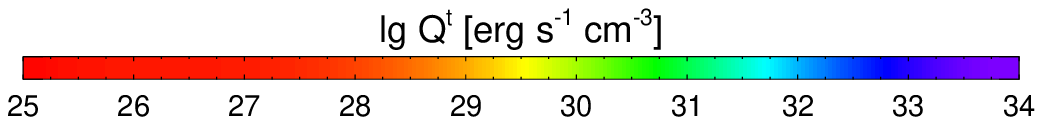}
  \end{center}
  \caption{Local annihilation rates for neutrinosphere geometries calculated from four equilibrium
           torus models (see Table \ref{tab:equilibrium}). The visualization method is the same as
           in Figs.\,\ref{fig:field_idealized1} and \ref{fig:field_idealized2}, but here the
           neutrinospheres for $\nu_{\rm e}$ (blue, long dashes) and $\bar{\nu}_{\rm e}$ (red, short
           dashes) do not coincide. Note that the accretion tori of the four models shown here
           rotate, however the spatial components of the annihilation rate 4-vector perpendicular to
           the displayed $x-z$-plane are very small and visible only in the close vicinity of the
           black hole.
           {\em See the electronic edition for a color version of the figure.}}
  \label{fig:field_equilibrium}
 \end{figure*}

\subsubsection{Influence of the angular momentum of the BH}
 \label{subsub:results_BH_rotation}

 The dependence of the energy deposition rate on the angular momentum of the central BH is studied
 using model DA1, which consists of a disk surrounding a maximally rotating Kerr BH ($a=1$; Table
 \ref{tab:idealized}; Fig.\,\ref{fig:field_idealized2}). Compared to the corresponding non-rotating
 model D ($a=0$), we do not find an enhancement of the energy deposition rate, independent of the
 region of integration and of the location of the observer (Table \ref{tab:comparison_idealized}).
 This result differs from that of \cite{2003.Miller}, who obtained a factor of $\sim 2$ larger
 amount of deposited energy when comparing thin accretion disks orbiting a maximally rotating and a
 non-rotating BH, respectively. This discrepancy arises because we set $r_{\rm in} = 6\,M$ both in
 model D and DA1, in order to just test the influence of BH rotation. In contrast, the disks
 considered by \citeauthor{2003.Miller} extended from a fixed outer radius $r_{\rm out}= 10\,M$ in
 towards the BH, i.e. to the innermost stable circular orbit. This results in disks with smaller
 values of $r_{\rm in}$ as $a \rightarrow 1$ than in our study, and hence, as we have shown in the
 last Section, an increased energy deposition rate is expected. Table \ref{tab:idealized} also
 contains the two models A.5 and A1, which are equal to our second reference model REF ($a=0$, and
 $M_{\rm BH} = 3 \Msun$, in contrast to $M_{\rm BH} = 2 \Msun$ for model D), except that the black
 hole rotates with $a=0.5$ and $a=1$, respectively. We find that for every considered BH mass the
 value of $\dot{E}_{\nu \bar\nu}^{{\rm up}, \infty}$ depends only weakly on the value of $a$.

\subsubsection{Influence of the accretion disk rotation}
 \label{subsub:results_disk_rotation}

 In all idealized models discussed up to now with the shape of an infinitely thin disk, a torus, or
 a sphere, respectively (Table \ref{tab:idealized}), the rotation of the neutrinosphere was
 neglected in calculating the \nnb-annihilation. We now address the consequences of relativistic
 aberration connected with such rotation on the energy deposition rate. Our idealized rotating disks
 are assumed to have a uniform Lagrangian angular momentum $l$. In Model DL6.5 the disk is rapidly
 rotating with the super-Keplerian value $l=6.5\,M$ (corresponding to a cgs value of the specific
 angular momentum of $l=5.7\cdot10^{16}\,{\rm cm}^2\,{\rm s}^{-1}$ for the $2\,M_{\odot}$ BH), but
 otherwise the same parameters as in the non-rotating model D ($l=0$) are used. For an observer at
 rest, the (anti)neutrinos are preferentially emitted tangentially to the disk in the direction of
 rotation. This behavior results in the existence of a non-vanishing $Q^{\phi}$ component and is
 reflected by the increased size of the filled circles in the vicinity of the disk in Fig.\,
 \ref{fig:field_idealized2}. Consequently, neutrinos and antineutrinos collide less frequently with
 large relative angles, and in the `tot' region of model DL6.5 less energy is deposited than in the
 non-rotating disk model D ($x_{\nu \bar\nu}^{\rm tot} < 1$; Table \ref{tab:comparison_idealized}).
 However, if we consider the `up' region, this effect is compensated by the slightly increased size
 of the `up' region in model DL6.5 ($x_{\nu \bar\nu}^{\rm up} \approx 1$), especially in the
 vicinity of the accretion disk (in model DL6.5 the white line crosses the equatorial plane at a
 smaller radius than in model D, see, Figs.\,\ref{fig:field_idealized1} and
 \ref{fig:field_idealized2}). The results also hold for an observer at infinity. Hence, independent
 of the choice of the observer, the influence of the disk rotation on the integrated energy
 deposition rate in the `up' region is small, which justifies the remarks of
 Sect.\,\ref{sub:results_idealized} regarding the choice of non-rotating reference models. This is
 also true for intermediate values of $l$ (see Table \ref{tab:idealized}), i.e. in case of the
 sub-Keplerian rotation of model L2.5, models DL4 and L4, whose accretion disk approximately rotates
 at Keplerian speed, and the super-Keplerian model L5 (the three models L2.4, L4, and L5 are not
 based on model D, but on model REF). Also in case of a toroidal neutrinosphere instead of a disk,
 the effect of the neutrinosphere rotation on the total annihilation rate measured by an observer at
 infinity is only small (see the values of $\dot{E}_{\nu\bar{\nu}}^{{\rm up}, \infty}$ for models T
 and TL4 in Table \ref{tab:idealized}).

 We have also studied the influence of the neutrinosphere rotation in case of Newtonian models. For
 this purpose, we have calculated models DL4N and TL4N (Table \ref{tab:idealized}). When GR effects
 are ignored we again find that the neutrinosphere rotation does not have an important impact on the
 total annihilation rate measured at infinity (compare $\dot{E}_{\nu\bar{\nu}}^{{\rm up}, \infty}$
 for models DL4N and DN, and TL4N and TN in Table \ref{tab:idealized}).

 Despite the weak dependence of the integrated values of the energy deposition rate in the `up'
 region on the specific angular momentum of the disk, the spatial distribution of the energy
 deposition rate changes significantly (compare the plots of models D, DL4, and DL6.5 in
 Figs.\,\ref{fig:field_idealized1} and \ref{fig:field_idealized2}). The faster the disk rotation is,
 the larger are the spatial components of annihilation rate 4-vector tangential to the disk. In
 addition to that, the energy deposited in the vicinity of the symmetry axis decreases, which is
 especially visible in model DL6.5.

 For the non-rotating model REF and the two rotating models L2.5 and L5 the left panel of Fig.\,
 \ref{fig:angular} shows the energy $\dif\dot{E}_{\nu\bar{\nu}}^{{\rm par},\infty}/\dif\theta$
 arriving at a sphere of radius $200\,M$ per unit of time and per unit of Boyer-Lindquist polar
 angle $\theta$. Numerically, these results are obtained by closely following the approach of
 \cite{2003.Miller}. We consider all points of our numerical grid and for each grid point we
 transport the calculated annihilation rate 4-vector $Q^{\alpha}$ parallel to itself, i.e. we
 follow geodesics in the direction of the 4-vector, starting at the annihilation point. Excluding
 the fraction of parallelly transported 4-vectors that eventually hit the black hole or the
 neutrinosphere, the remaining 4-vectors $Q^{\alpha}$ are transported until they hit the sphere of
 radius $200\,M$ (centered in the system origin). For each such 4-vector, we take the initial energy
 component $Q^{t}$, i.e. before the parallel transport, and multiply it by the factor $2\pi\Delta
 r\Delta\theta\sqrt{ -\det\left( g_{\alpha\beta} \right) }$, in analogy to Eqs.\,(\ref{edotitot})
 and (\ref{edotiup}). This way we take into account the 3-volume and gravitational redshift.
 Finally, depending on the $\theta$-angle of the 4-vector $Q^{\alpha}$ arriving at the sphere, we
 create the plots of Fig.\,\ref{fig:angular} by using 20 equally spaced $\theta$-bins.

 In case of the rotating thin disk models L2.5 and L5 (see Fig.\,\ref{fig:angular}) we obtain
 results very similar to those of \cite{2003.Miller}, namely the dominant contribution to $\dif
 \dot{E}_{\nu\bar{\nu}}^{{\rm par},\infty}/\dif\theta$ is found to be off-axis and scaling their
 results to cgs units they quantitatively agree with ours (compare Fig.\,8 in their work with the
 left panel of our Fig.\,\ref{fig:angular}). However, the position of the peak strongly depends on
 the disk rotation, and for non-rotating models the bulk of the energy is deposited in the vicinity
 of the system axis. In addition to the disk models REF, L2.5, and L5, we have also considered the
 non-rotating torus model AB and rotating idealized torus models (not shown in this work), and we
 find the same general behaviour as in case of the disk models.

\subsubsection{Efficiency}

 The efficiency of converting radiated neutrino energy into ${\rm e}^+{\rm e}^-$-pairs by
 \nnb-annihilation is given by the quantity $q_{\nu\bar\nu}^{{\rm up},\infty} \equiv \dot{E}_{\nu
 \bar\nu}^{{\rm up},\infty} / L^{\infty}_\nu$, where $L^{\infty}_\nu$ is the sum of the neutrino and
 antineutrino luminosities at infinity (see Table \ref{tab:idealized}). Values as large as a few
 tenths of a percent can be reached for $q_{\nu\bar\nu}^{{\rm up}, \infty}$. Neutrinospheres having
 the geometry of a thin disk are the most efficient ones in converting neutrino energy into ${\rm
 e}^+{\rm e}^-$-pairs, while neutrinospheres of toroidal shape have the smallest efficiency (see the
 models D, T, and S in Table \ref{tab:idealized}). If GR effects are not taken into account in
 calculating the energy deposition rate by \nnb-annihilation (see models DN, TN, and SN), thin disks
 are still most efficient, however, now the spherically shaped neutrinosphere is least efficient.
 Higher disk luminosities increase the deposition efficiency (see the model REF and all models below
 in Table \ref{tab:idealized}). The efficiency monotonically depends on the inner radius $r_{\rm
 in}$ of accretion disks (see models REF, RI4.5, and RI3 in Table \ref{tab:idealized}). When
 $r_{\rm in}$ is reduced, the neutrinosphere moves closer to the event horizon of the BH. Although,
 this way the neutrino luminosity measured by an observer at the locations of \nnb-annihilation
 rises (not shown in Table \ref{tab:idealized}), the neutrino luminosity at infinity $L_{\nu}^
 {\infty}$ nearly remains the same due to the increased gravitational redshift. Since the energy
 deposited at infinity $\dot{E}_{\nu\bar\nu}^{{\rm up},\infty}$ increases, the efficiency becomes
 larger for smaller values of the inner disk radius $r_{\rm in}$. An analogous effect can also be
 seen in the torus model TEMP (see Table \ref{tab:idealized}), where compared to the torus model AB
 the accretion torus and Newtonian luminosity $L_{\nu}^{{\rm N}}$ are the same, but the radiating
 region is closer to the BH. In contrast, up to two significant figures the efficiency exhibits no
 dependence on the angular momentum of the black hole (see, \eg model A.5 compared to model REF) and
 of the disk (see models L2.5, L4, and L5 in Table \ref{tab:idealized}).

\subsection{Equilibrium models}
 \label{sub:results_equilibrium}

 \begin{figure}
  \includegraphics[width=0.5\textwidth]{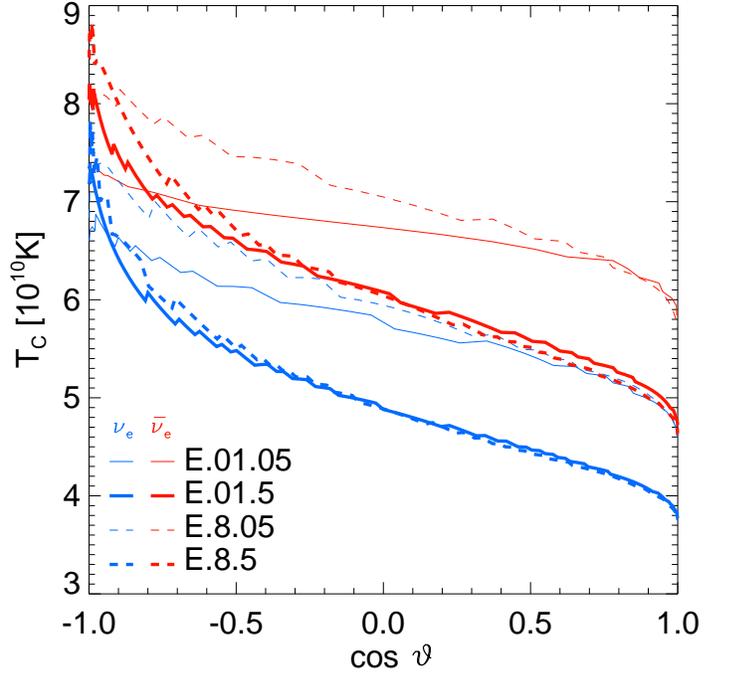}
  \caption{Temperature profiles of the neutrinospheres of the models in Fig.\,
           \ref{fig:field_equilibrium} in the comoving frame versus the angle $\vartheta$, which is
           defined as follows: If $r_{\rm c}$ is the arithmetic mean of the innermost ($r_{\rm in}$)
           and outermost ($r_{\rm out}$) radius where the neutrinosphere intersects the equatorial
           plane, then given a point $C$ on the neutrinosphere the angle $\vartheta$ is enclosed by
           the legs $\overline{r_{\rm c} r_{\rm out}}$ and $\overline{r_{\rm c}C}$, respectively.
           The blue (red) lines denote the temperature of the neutrinosphere (antineutrinosphere).
           {\em See the electronic edition for a color version of the figure.}}
  \label{fig:temperatures}
 \end{figure}

 \begin{table}
  \tabcolsep=3.2mm
  \caption{Comparison of the equilibrium torus models. The table is structured like Table
           \ref{tab:comparison_idealized}, but the models shown here are models listed in Table
           \ref{tab:equilibrium}.
  \label{tab:comparison_equilibrium}}
  \begin{center}
   \begin{tabular}{cc|cccc}
    \hline\hline &&&&& \\*[-0.35cm]
     M1 & M2 &
     $x_{\nu\bar\nu}^{\rm tot}$ &
     $x_{\nu\bar\nu}^{\rm up}$ &
     $x_{\nu\bar\nu}^{{\rm tot},\infty}$ &
     $x_{\nu\bar\nu}^{{\rm up},\infty} $ \\*[0.05cm]
    \hline
     E.01.5  & E.01.05 & 7.7 & 7.5 & 7.8 & 7.5 \\
     E.8.5   & E.8.05  & 3.4 & 4.0 & 3.9 & 4.0 \\
     E.8.05  & E.01.05 & 3.5 & 2.5 & 3.2 & 2.6 \\
     E.8.5   & E.01.5  & 1.6 & 1.3 & 1.6 & 1.4 \\
    \hline
     E1.1d   & E.01.1  & 1.0 & 1.0 & 1.1 & 1.0 \\
     E1.1    & E.01.1d & 1.1 & 1.0 & 1.1 & 1.0 \\
     E.01.1d & E.01.1  & 2.1 & 1.7 & 1.9 & 1.6 \\
     E1.1    & E1.1d   & 2.5 & 1.7 & 2.1 & 1.6 \\
    \hline
     E.01.25 & E.01.05 & 5.5 & 5.3 & 5.4 & 5.3 \\
     E.01.1  & E.01.1N & 2.3 & 1.3 & 1.7 & 1.2 \\ \hline
   \end{tabular}
  \end{center}
 \end{table}

 Next we discuss the \nnb-annihilation rate in the vicinity of the neutrinospheres of equilibrium
 accretion tori computed as described in Sect.\,\ref{sub:theory_neutrinosphere}. In all of our
 equilibrium torus models we keep $s_{\gamma}$, $r_{\rm in}$, and $Y_{\rm e}$ fixed, and therefore
 we will consider only the influence of the three parameters $(M,a,m_{\rm tor})$ in the following
 paragraphs.

 In contrast to the idealized models, the neutrinosphere and the antineutrinosphere do not coincide
 in the equilibrium models (Fig.\,\ref{fig:field_equilibrium}), because there is a larger number
 density of free neutrons than free protons in the equilibrium tori. Hence, the opacity for
 neutrinos and antineutrinos differs, and consequently the locations of the neutrinosphere and
 antineutrinosphere, too. Since the matter in the equilibrium tori is neutron rich, the
 antineutrinosphere is always completely interior to the neutrinosphere. On average it also has a
 higher temperature (Figs.\,\ref{fig:field_equilibrium} and \ref{fig:temperatures}), i.e. $L_{\bar
 \nu} \geq L_{\nu}$. Neither the neutrinosphere nor the antineutrinosphere are isothermal as in case
 of the idealized models (Fig.\,\ref{fig:temperatures}). Moreover, the torus temperature now depends
 on the BH spin and increases, the closer the torus is located around the BH. This effect is ignored
 in the idealized models.

 The energy deposited in the region between the neutrinosphere and the antineutrinosphere cannot be
 computed by our approach (hence the corresponding region is white in Fig.\,
 \ref{fig:field_equilibrium}), because in that region transport physics matters, i.e. neutrinos do
 not propagate freely.

 In order to analyze the influence of the torus mass $m_{\rm tor}$, let us consider the equilibrium
 models (Table \ref{tab:equilibrium}) E.01.05, E.01.25 and E.01.5, all of which orbit a slowly
 rotating BH with $a=0.01$, and models E.8.05 and E.8.5, which gird a rapidly rotating BH with
 $a=0.8$. From Table \ref{tab:comparison_equilibrium} we infer (i) that a larger torus mass strongly
 increases the \nnb-energy deposition rate, because of a higher neutrino luminosity, and (ii) that
 this effect is stronger for lower values of $a$. The increase of the energy deposition rate is
 almost linear with $m_{\rm tor}$ for high values of $a$, but grows like $m_{\rm tor}^{0.6}$ for
 $a=0.01$. Thus, for a distant observer the energy released in the `up' region can be roughly fitted
 by $\dot{E}_{\nu \bar\nu}^{{\rm up}, \infty} \propto \dot{E}_{\nu\bar\nu}^{{\rm up},\infty}({\rm
 E.01.05}) \cdot (m_{\rm tor} / M_{\sun})^{0.4a+0.6}$.

 On the other hand, comparing models with the same torus mass but with BHs that have different $a$,
 e.g. models E.01.05 and E.8.05, shows that larger values of $a$ lead to a larger energy deposition
 ($x_{\nu\bar\nu}^{{\rm up},\infty} = 2.6$; Table \ref{tab:comparison_equilibrium}). The effect of
 $a$ on the deposition rate is about a factor of 2 smaller for higher torus masses (compare models
 E.8.5 and E.01.5 in Table \ref{tab:comparison_equilibrium}).

 What causes the increased energy deposition rate in case of faster rotating BHs? There are two
 possibilities: the rotation rate parameter $a$ influences (i) the geometry of the accretion torus
 and hence the neutrinosphere, and (ii) it affects the (anti)neutrino trajectories through frame
 dragging. In order to determine the importance of both possibilities, we consider artificial test
 models where we construct the torus model for a rotating BH but then disregard the BH rotation for
 computing the neutrino geodesics, or vice versa. This means effectively that we consider two
 different rotation parameters for the same model. The first one, $a_{\rm tor}$, is used to
 calculate the equilibrium tori, and hence the neutrinospheres. The second one, $a$, is used to
 calculate the GR effects on the propagation and annihilation of neutrinos and antineutrinos. Note
 that the smaller of both values is used to set the inner edge of the computational grid, which is
 thus located slightly outside the larger of both horizons. Such models are actually inconsistent,
 because a change in the BH spin automatically translates into a structural change of the accretion
 torus. We nevertheless consider them in order to be able to discriminate the influence of the BH
 rotation on the \nnb-annihilation through its effects on the neutrino trajectories from those
 effects that result from changes of the torus properties around rotating or non-rotating BHs.

 For each of the two parameters, $a$ and $a_{\rm tor}$, we consider two extreme values, which leads
 to the four models E.01.1, E.01.1d, E1.1d, and E1.1, respectively (Table \ref{tab:equilibrium}). As
 in case of the idealized models (Sect.\,\ref{sub:results_idealized}), increasing the BH's rotation
 rate from $a = 0.01$ to $a=1$ does not yield any appreciable increase of the energy deposition rate
 ($x_{\nu \bar\nu}^{{\rm up}, \infty} \approx 1$) for both values of $a_{\rm tor}$ (Table
 \ref{tab:comparison_equilibrium}). However, a substantial increase ($\sim 60\%$) of the energy
 released in the `up' region (as seen by a distant observer) is found when increasing $a_{\rm tor}$
 from 0.01 to 1. Thus, we can unambiguously conclude that the increase of the dimensionless angular
 momentum of the BH yields a larger energy deposition by \nnb-annihilation essentially exclusively
 because it allows for a smaller innermost radius of the equilibrium accretion torus.

 Concerning the influence of the GR effects, a comparison of model E.01.1 with its Newtonian
 counterpart E.01.1N (Table \ref{tab:equilibrium}) shows that GR effects on the neutrino propagation
 increase $\dot{E}_{\nu\bar\nu}^{{\rm up}, \infty}$ in case of the equilibrium models only by
 $\sim 20\%$. This enhancement of the energy deposition rate is similar to that obtained for the
 idealized torus models (see Sect.\,\ref{sub:results_idealized}).

 In analogy to the idealized models, we also computed the energy deposition rate distribution on a
 sphere of radius $200\,M$ per polar angle $\theta$ for equilibrium models (see right panel of Fig.
 \,\ref{fig:angular}). However, in contrast to the idealized models, the bulk of the energy is
 always deposited near the system axis. Hence, the result of \cite{2003.Miller}, i.e. that the main
 contribution to the energy deposition rate comes from the off-axis region, is valid only for
 idealized models (see Sect.\,\ref{subsub:results_disk_rotation}), but in equilibrium torus models
 the energy deposition along the symmetry axis is clearly dominant.

 The efficiency $q_{\nu \bar\nu}^{{\rm up}, \infty}$ increases for our equilibrium models both with
 the torus mass and with $a$, reaching values of 1-3 per cent (Table \ref{tab:equilibrium}) in
 agreement with the results of \cite{1996.Jaroszynski} for his models with specific entropies $\gsim
 8$, and with \cite{2004.Setiawan} for their models with the largest $\alpha$-viscosity. The
 increase with $a$ is a structural effect, because for larger values of $a$ the torus is on average
 closer to the rotation axis and hotter, i.e. the neutrino luminosity is higher. Therefore the
 neutrino density is higher near the system axis and hence a larger fraction of neutrinos and
 antineutrinos annihilate with each other. The efficiencies of the equilibrium models are $\sim 10$
 times larger than those of the idealized models, because both the temperatures (Fig.\,
 \ref{fig:temperatures}) and the surface areas of their neutrinospheres are larger.

\section{Conclusions}
 \label{sec:conclusions}

 The main goal of this paper is to study generic properties of the process of energy deposition by
 neutrino-antineutrino (\nnb) annihilation in the vicinity of systems consisting of a central
 stellar-mass black hole (BH) and an accretion disk or torus surrounding it. Assuming that the
 thermodynamic conditions in the accretion flow are such that a copious flux of \nnb-pairs can be
 produced, we have performed a systematic parameter study of the influence of the mass and
 dimensionless angular momentum parameter of the BH, of the shape and thermal properties of the
 neutrinosphere, and of the importance of different general relativistic (GR) effects on the amount
 of energy released by \nnb-annihilation.

 On the one hand we considered idealized models having an isothermal neutrinosphere of prescribed
 temperature and geometry, and on the other hand non-selfgravitating, axisymmetric equilibrium tori
 bound to a central BH of given properties. In the latter models the neutrinospheres of neutrinos
 and antineutrinos do not coincide, because they are computed for opacities of $\nu_{\rm e}$ and
 $\bar{\nu}_{\rm e}$ that differ due to absorptions on free neutrons or free protons, respectively
 (similar to the work of \citealp{1993.Jaroszynski,1996.Jaroszynski}). Using both sets of models we
 numerically calculated the annihilation rate energy-momentum 4-vector. For this purpose, we
 constructed the local neutrino distribution by ray-tracing neutrino trajectories in a Kerr
 space-time using GR geodesics. Our study is a generalization of the work of \cite{2003.Miller} to
 axisymmetric neutrinospheres of different shapes, which are not necessarily spatially coincident
 for $\nu_{\rm e}$ and $\bar{\nu}_{\rm e}$.

 We considered three different neutrinosphere geometries in our set of idealized models covering
 cases that may be encountered in astrophysical systems: infinitely thin disks, tori, and spheres.
 Infinitely thin disks are mathematical idealizations. However, they are investigated as the
 limiting case of very oblate tori. For this reason they have been widely used in the literature. By
 comparing with corresponding Newtonian models, where the influence of the gravitational field on
 the neutrino propagation is neglected, we have found that for an observer at infinity GR effects
 enhance the energy deposition rate by a factor of 2 in the non-rotating, thin disk case in
 agreement with \cite{2001.Asano}. In the other two cases the influence of GR effects enhances the
 annihilation rate only by $\approx 25\%$.

 Independent of whether GR effects are included, the energy deposition rate that may lead to the
 acceleration of relativistic outflow is largest for infinitely thin disks. For more prolate tori
 the energy deposition rate drops in comparison with a disk that has the same neutrino luminosity,
 because a sizeable fraction of the energy deposited by \nnb-annihilation is swallowed by the BH.
 Spherical neutrinospheres are the least favorable geometry for a large total annihilation rate.
 Accretion disks encountered in gamma-ray burst (GRB) scenarios are likely to be geometrically
 thick, in which case the geometry of their neutrinospheres is torus-like. For such a situation
 neglecting GR effects in the evaluation of \nnb-annihilation is a good approximation.

 We also analyzed the influence of the dimensionless rotation parameter $a$ of the BH, and we
 confirm the findings of \cite{2003.Miller} that modifying $a$ without changing the innermost disk
 radius (\ie leaving the thin disk geometry unchanged) has no effect on the amount of released
 \nnb-annihilation energy. The independence of the results of the BH rotation holds both for
 idealized models with toroidal neutrinospheres and for equilibrium accretion torus models. However,
 in a consistent accretion disk model, the innermost radius of the disk (close to the innermost
 stable circular orbit) will shrink as $a\rightarrow 1$. We have found that a smaller innermost disk
 radius leads to a substantial increase of the energy deposition rate, even when the local Newtonian
 (\ie GR effects disregarded) neutrino luminosity is the same. This can be understood by the higher
 neutrino density in the vicinity of a more compact disk and was also found by \cite{2003.Miller}.

 Depending on the mass of the accretion torus, models containing a maximally rotating BH ($a=1$) can
 release roughly a factor of 2 more energy (as measured by a distant observer) by \nnb-annihilation
 than models involving non-rotating BHs. For a small accretion disk mass ($\sim 0.05 \Msun$) the
 energy release in case of a rotating ($a=0.8$) BH is more than 2.5 times larger than in case of a
 non-rotating BH. The difference is smaller for larger torus masses. Considering that low-mass
 accretion tori rotating around a BH with intermediate values of the dimensionless angular momentum
 parameter ($a \sim 0.6-0.8$) may be typical products of mergers of compact objects (\eg
 \citealp{1997.Ruffert,2004.Setiawan}), we have demonstrated that all other GR effects are of
 moderate importance when calculating the amount of energy released in such systems by
 \nnb-annihilation. However, we point out that the spatial distribution of the released energy
 exhibits very important differences between Newtonian and GR models: Close to the rotation axis and
 close to the bounding of the `up' region, GR effects can enhance the local energy deposition rate
 by a factor of $\sim 10$. Thus, a relativistic jet driven by the energy release in a region close
 to the stagnation surface (which will form in the vicinity of the boundary of the `up' region)
 should receive a much larger energy input due to GR effects. The question whether this difference
 in the spatial distribution of the energy deposition is more favorable for producing
 ultrarelativistic jets can not be addressed by the present work, but requires time-dependent
 hydrodynamic jet simulations including a detailed description of \nnb-annihilation (see below).

 In our set of idealized models, a thin disk (sphere) is the most (least) favorable geometry to
 convert neutrino energy into ${\rm e}^+{\rm e}^-$-pairs by \nnb-annihilation. Our idealized models
 yield efficiencies $q_{\nu\bar\nu}^{{\rm up},\infty}$ of only a few tenths of a per cent. Larger
 efficiencies, of the order of several percent, have been obtained for our more realistic
 equilibrium models. These findings are in the ball-park of the results obtained by
 \cite{1996.Jaroszynski} (for models with specific entropies $\gsim 8$ and \cite{2004.Setiawan}; for
 models with the largest $\alpha$-viscosity).

 \cite{2003.Miller} have pointed out that the main contribution to the energy-momentum deposition
 rate comes from large polar angles (particularly from regions above the neutrinospheres) and not
 from regions near the symmetry axis. This behavior, however, results from their restriction to
 infinitely thin disk models, which are mathematically idealized cases. According to our studies,
 only such idealized models show this strong enhancement of the annihilation rate towards the
 equatorial plane, while this effect is absent in the more realistic equilibrium torus models.
 Hence, this finding of \cite{2003.Miller} appears to be of little relevance for the amount of
 energy that is available to produce a GRB. In any realistic situation the accretion disk/torus will
 be inflated vertically, i.e. the baryon density will be highest in the equatorial plane and
 decrease in perpendicular direction, the density scale height of the baryon distribution being
 larger for large disk masses and high disk temperatures.

 As demonstrated in \cite{2005.Aloy}, who considered similar equilibrium tori as initial models for
 their hydrodynamic simulations of relativistic jets from BH-torus systems, time-dependent numerical
 simulations are needed in order to determine the feedback of \nnb-energy deposition on the torus
 structure, which, in its turn, can yield a number of highly non-linear relativistic hydrodynamic
 effects on the outflowing jet \citep{2006.Aloy}. This also decides about which fraction of the
 energy deposited by \nnb-annihilation is finally useful for accelerating an ultrarelativistic
 fireball powering a GRB event. This fraction can be different from the annihilation efficiency
 $q_{\nu\bar\nu}^{{\rm up},\infty}$ defined in the present work. Our equilibrium models produce
 \nnb-annihilation rates $\dot{E}_{\nu\bar\nu}^{{\rm up},\infty} \gsim 10^{52}$\ergsec\, for torus
 masses $m_{\rm tor} \gsim 0.1\Msun$. These figures agree within a factor of 2--4 with the results
 of dynamical simulations (\citealp{2005.Setiawan}) and they are about one order of magnitude higher
 than the ones quoted in \cite{1996.Jaroszynski} for his models with specific entropies per baryon
 in the range of 8 to 10. The difference compared to \citeauthor{1996.Jaroszynski}'s estimates is
 probably due to the differences in the temperature distribution on the neutrinosphere caused by our
 somewhat different way of constructing the accretion tori. The neutrino luminosity is very
 sensitive to the neutrinospheric temperature, and a $\sim 30 \%$ larger value of this temperature
 can account for a factor of more than 10 larger total \nnb-annihilation rate. Considering a certain
 accretion disk mass and its corresponding neutrino luminosity and annihilation efficiency means an
 upper bound of the physically likely situation, in which the neutrino luminosity decreases with
 time as the mass of the disk decreases. However, we remark that the time evolution of the
 \nnb-annihilation rate and the annihilation efficiency can be non-monotonic and does not
 necessarily decrease with time as demonstrated by \cite{2005.Setiawan}.

 When attempting to link our present results to GRB observations, we must therefore be aware of the
 restrictions stated in the previous paragraph and in particular of the fact that any reliable
 estimate of the fraction of the \nnb-annihilation energy that drives an ultrarelativistic fireball
 and the determination of the collimation of such outflow requires hydrodynamic simulations (see
 also \citealp{2006.Janka}). So we conclude with caution that the energy release by
 \nnb-annihilation in some of our models (especially those having the most massive tori or large
 values of $a$) could be sufficient even to fuel the most distant and most powerful short GRB
 discovered so far (GRB060121, $E_{\gamma} \approx 3 \cdot 10^{51}\,$erg; \citealp{2006.Postigo})
 without the need of invoking either ultra-intense magnetic fields ($B>10^{16}\,$G) or a different
 progenitor class.

 Finally, we point out that \citeauthor{2005.Aloy} employed a simple fit with time-independent
 geometry to describe the energy deposition by \nnb-annihilation above the poles of a stellar-mass
 BH in their relativistic hydrodynamic jet simulations. Preferably, future numerical simulations
 should include a time-dependent treatment of the energy deposition by \nnb-annihilation using the
 refined methods employed in this work.

 \begin{acknowledgements}
  MAA is a Ram\'on y Cajal Fellow of the Spanish Ministry of Education and Science. He also
  acknowledges partial support from the Spanish Ministry of Education and Science
  (AYA2004-08067-C03-C01). This work has been supported in part by the
  Sonderforschungsbereich-Transregio 7 `Gravitationswellenastronomie' of the German Science
  Foundation (DFG).
 \end{acknowledgements}

 \appendix

\section{Some technical details of the calculation of the annihilation rate}
 \label{sec:appendix_maths}

 Here we describe in detail the evaluation of Eq.\,(\ref{eq:1}), which is a natural generalization
 of the formula for the energy component $Q_{i}^{t}$ given in \cite{1997.Ruffert}.

 We express the neutrino 3-momentum $\vec{p}$ (analogous expressions hold for the antineutrino
 quantities which are represented by primed variables) in terms of the energy $E = \left|\vec{p}
 \right|$ and the unit vector in the momentum direction $\vec{n} = \vec{p}/E$. The distribution
 function is defined as
 \[
  f\equiv f\left(t,\vec{x},E,\vec{n}\right):=\frac{h^3}{g}\frac{\dif
  N}{\dif V_x\dif V_p},
 \]
 where $h$ is Planck's constant, and $g$ is the statistical weight ($g=1$ for neutrinos and
 antineutrinos). Using $\dif^{3}p = E^2\dif E \dif\Omega$ and defining $\theta$ as the angle between
 the directions of propagation of the neutrino and antineutrino, Eq.\,(\ref{eq:1}) in full detail
 reads
 \begin{eqnarray*}
  Q_i^\alpha
   & =  & \frac{1}{4} \frac{\sigma_0}{m_{\rm e}^2 h^6}\\
   &    & \left\{ \frac{\left(C_1+C_2\right)_{\nu_i\bar{\nu}_i}}{3}
          \int_0^\infty \dif E \int_0^\infty \dif E'
          \left(p^\alpha+p'^\alpha\right)E^3 E'^3\right.\\
   &    & \;\;\;\;\;\;\;\;\;\;\;\;\;\;\;\;\;\;\;\;\;\;\;\;\;
          \oint_{4\pi} \dif\Omega \oint_{4\pi} \dif\Omega'
          \left( 1-\cos\theta \right)^2
          f_{\nu_i} f_{\bar{\nu}_i} + \\
   &    & \;\;\;\;\;\;\;\;\,
          C_{3,\nu_i \bar{\nu}_i} m_{\rm e}^2
          \int_0^\infty \dif E \int_0^\infty
          \dif E'\left(p^\alpha+p'^\alpha\right)E^2 E'^2\\
   &    & \;\;\;\;\;\;\;\;\;\;\;\;\;\;\;\;\;\;\;\;\;\;\,
          \left.\oint_{4\pi} \dif\Omega \oint_{4\pi} \dif\Omega'
          \left(1-\cos\theta\right) f_{\nu_i} f_{\bar{\nu}_i} \right\},
 \end{eqnarray*}
 where $p^\alpha = \left(E,\vec{p}\right)$, $\sigma_0 = 1.76\cdot 10^{-44}\,$cm$^2$ is the weak
 interaction cross section, $m_{\rm e}$ the electron mass, and finally
 \begin{eqnarray*}
  \left(C_1+C_2\right)_{\nu_e\bar{\nu}_e} \approx 2.34,
  & \;\;\;\;\;\;\;\;\;\; &
  \left( C_1+C_2 \right)_{\nu_x\bar{\nu}_x} \approx 0.50,\\
  C_{3,\nu_e\bar{\nu}_e} \approx 1.06,\,
  & & \;\;\;\;\;\;\;\;\;\;\;
  C_{3,\nu_x\bar{\nu}_x} \approx -0.16,
 \end{eqnarray*}
 with $x\in\left\{ \mu,\tau\right\}$.

 In order to perform the ray-tracing of the neutrino trajectories a definition of the base vectors
 spanning the local observer frame is required. Our choice is the same as that of
 \cite{2003.Miller}, i.e.
 \begin{eqnarray*}
  &  & \left(\boldsymbol{e}_t{}^\alpha\right)
      =\frac{1}{\sqrt{g_{tt}}}\left(1,0,0,0\right),
       \left(\boldsymbol{e}_r{}^{\alpha}\right)
      =\frac{1}{\sqrt{-g_{rr}}}\left(0,1,0,0\right),\\
  &  & \left(\boldsymbol{e}_\theta{}^\alpha\right)
      =\frac{1}{\sqrt{-g_{\theta\theta}}}\left(0,0,1,0\right),\\
  &  &
      \left(\boldsymbol{e}_\phi{}^\alpha\right)
     =\sqrt{\frac{g_{tt}}{g_{t\phi}^2-g_{\phi\phi}g_{tt}}}
      \left(-\frac{g_{t\phi}}{g_{tt}},0,0,1\right),
 \end{eqnarray*}
 where the metric coefficients are given in Eq.\,(\ref{kerrmetric}). Note that
 \citeauthor{2003.Miller} used the same signature as \cite{1973.Misner} and that the selected base
 is orthonormal, i.e. $\boldsymbol{e}_{\alpha} \cdot \boldsymbol{e}_{\beta} = \eta_{\alpha\beta}$,
 where $\eta_{\alpha\beta}$ is the Minkowski metric.

\section{Convergence Tests}
 \label{sec:appendix_convergence}

\begin{figure}
  \includegraphics[width=0.5\textwidth]{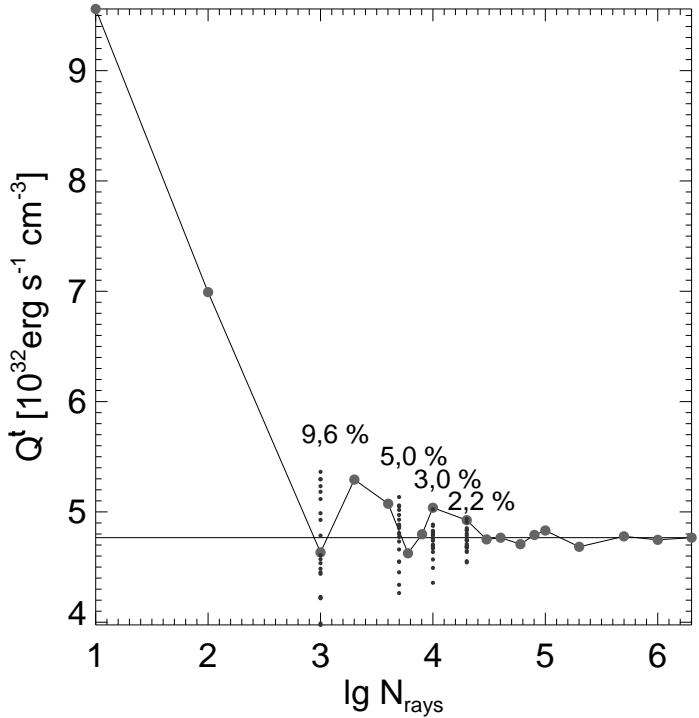}
  \caption{Convergence behaviour of the energy component $Q^t$ of the annihilation rate 4-vector for
           a $M = 1\,\Msun$, $a=0$ black hole. The neutrinosphere (for both $\nu_{\rm e}$ and $\bar
           \nu_{\rm e}$) is considered to be a non-rotating isothermal thin disk of temperature $T =
           10^{11}\,$K extending from $r_{\rm in} = 3.5M$ to $r_{\rm out} = 5.5M$. The rate is
           computed at the point ($r = 3.4M$, $\theta = \frac{\pi}{4}$) using different numbers $N_{
           \rm rays}$ of ray-tracing paths for neutrinos and antineutrinos. The horizontal line
           marks the asymptotic value of $Q^t$ as $N_{\rm rays} \rightarrow \infty$.
  \label{fig:convergence}}
 \end{figure}

 We performed a series of convergence tests to check the dependence of the energy component $Q^t$ of
 the annihilation rate 4-vector on the number $N_{\rm rays}$ of ray-tracing paths for neutrinos and
 antineutrinos. The results of this test are shown in Fig.\,\ref{fig:convergence}. The dots give the
 values of $Q^t$ at selected points for a representative BH-disk configuration. Successively
 increasing $N_{\rm rays}$ (gray filled circles connected by the black, solid line) we found that
 the value of $Q^t$ is converged to an accuracy better than $\sim 3\%$ for $N_{\rm rays} > 10^4$.
 When calculating $Q^t$ repeatedly for four selected values of $N_{\rm rays}$ the results scatter
 statistically due to the random procedure used for picking the initial direction of the ray-tracing
 paths for the (anti)neutrinos. The corresponding relative error is shown in Fig.\,
 \ref{fig:convergence}, too. Since all the simulations presented in this publication have been
 performed with $N_{\rm rays} = 20000$, the relative error of the calculated annihilation rates
 $Q^{t}$ is about $2\%$. This accuracy slightly varies depending on where the annihilation rate is
 calculated. Analogous tests were also performed for the spatial components of the annihilation rate
 4-vector, which show a similar convergence behaviour.

 A second series of convergence tests was computed, in order to check the dependence of the total
 annihilation rate on the grid resolution. In this context model REF, which was simulated on a
 computational grid of $N_r = 110$ and $N_\theta = 100$ points, was recalculated on a grid with
 twice as many grid points in each coordinate direction (\ie $N_{r} = 220$, and $N_\theta = 200$).
 Doubling the resolution the value of $\dot{E}_{\nu\bar \nu}^{{\rm tot,\infty}}$ defined in Eq.\,
 (\ref{edotitot}) changes from $6.19\cdot 10^{49}\,$erg\,s$^{-1}$ (coarse grid) to $6.21\cdot
 10^{49}\,$erg\,s$^{-1}$ (finer grid), which is a small difference of $0.3\%$.

 Finally, we investigated the dependence of the results on the size of the grid, i.e. on the value
 of the outer grid radius $r_{\rm g,out}$. Again using model REF, we increased the outer radius from
 its standard value $r_{\rm g,out}=48.75\,$km to $r'_{\rm g,out}=270\,$km. The resulting difference
 in $\dot{E}_{\nu\bar\nu}^{{\rm tot},\infty}$ is $2.5\cdot 10^{48}\,$erg\,s$^{-1}$, corresponding to
 a relative error of $\sim 4\%$. This error is also representative for the other models (note that
 there is negligible energy deposition for $r > r'_{\rm g,out}$).

 \bibliographystyle{aa}
 \bibliography     {6293}
\end{document}